\documentclass[a4paper,11pt]{article}
\pdfoutput=1 

\usepackage{jcappub} 
\usepackage[T1]{fontenc} 
\usepackage[utf8]{inputenc}

\usepackage{graphicx,amssymb,amsmath,amsthm,amsfonts,epsfig,times,bm}
\usepackage{aas_macros}
\usepackage{cleveref}

\usepackage{amsmath,amssymb}
\usepackage{tensor}
\usepackage{multirow}

\def\be{\begin{equation}}
\def\ee{\end{equation}}
\newcommand{\bea}{\begin{eqnarray}}
\newcommand{\eea}{\end{eqnarray}}

\title{\boldmath Rotating Axion Boson Stars}

\author[1]{Jorge F. M. Delgado,}\emailAdd{jorgedelgado@ua.pt}
\author[1]{Carlos A. R. Herdeiro,}\emailAdd{herdeiro@ua.pt}
\author[1]{Eugen Radu,}\emailAdd{eugen.radu@ua.pt}

\affiliation[1]{Departamento de Matemática da Universidade de Aveiro and 
Center for Research and Development in Mathematics and Applications -- CIDMA
Campus de Santiago, 3810-183 Aveiro, Portugal}

\abstract
{
We construct and study rotating axion boson stars (RABSs). These are the spinning generalisations of the spherical gravitating solitons recently introduced in~\cite{Guerra:2019srj}. RABSs are asymptotically flat, stationary, axially symmetric, everywhere regular solutions of the Einstein-Klein-Gordon theory, in the presence of a periodic scalar potential arising in models of axion-like particles. The potential is characterised by two parameters: the mass of the scalar field $m_a$ and the decay constant $f_a$. We present an overview of the solution space, for different values of  $f_a$, and analyse some of their phenomenological properties. For large decay constants the solutions become identical to the standard spinning mini boson stars. For small decay constants, on the other hand, the solutions develop distinctive features. In particular, we analyse their compactness, the emergence of ergoregions, light rings and the distribution of stable and unstable equatorial timelike circular orbits, including the innermost stable circular orbit. We also establish the occurrence of violations of the strong energy condition for physical observers, for some RABSs. We observe some analogy between RABSs and spinning gravitating $Q$-balls.
}

\begin{document}

\maketitle
\flushbottom

\section{Introduction}
Bosonic stars have attracted much interest in strong gravity research. Various reasons support this interest. Their original construction was motivated by the search of \textit{geons}, suggested by Wheeler~\cite{Wheeler:1955zz}. In modern terminology, this is what one may call self-gravitating solitons. Electrovacuum does not seem to allow solitons (see~\cite{Herdeiro:2019oqp} for an overview), but they exist in Einstein-Klein-Gordon theory. The original spherical scalar boson stars~\cite{Kaup:1968zz,Ruffini:1969qy}, obtained for a massive complex scalar field minimally coupled to Einstein's General Relativity, allowed many generalisations. These include the addition of several types of scalar self-interactions, $e.g.$~\cite{Colpi:1986ye,Kleihaus:2005me,Kleihaus:2007vk,Kleihaus:2015iea,Herdeiro:2015tia}, the inclusion of rotation~\cite{Schunck:1996he,Yoshida:1997qf}, see also $e.g.$~\cite{Herdeiro:2015tia,Grandclement:2014msa,Herdeiro:2019mbz}, the construction of vector analogues (Proca stars)~\cite{Brito:2015pxa,Herdeiro:2017fhv,Minamitsuji:2018kof}, including other asymptotics, $e.g.$~\cite{Astefanesei:2003qy,Duarte:2016lig}, multi-field configurations~\cite{Alcubierre:2018ahf}, alongside many others - see the reviews~\cite{Schunck:2003kk,Liebling:2012fv}.

Bosonic stars are, moreover, dynamically interesting solutions, in the sense some of the models are perturbatively stable~\cite{Gleiser:1988ih,Balakrishna:1997ej,Brito:2015pxa,Sanchis-Gual:2017bhw} and allow all sorts of dynamical studies, in particular of their dynamical formation~\cite{Seidel:1993zk,DiGiovanni:2018bvo} and of their evolution in binaries, producing gravitational waveforms of interest for ongoing LIGO/Virgo searches, $e.g.$~\cite{Palenzuela:2007dm,Bezares:2017mzk,Palenzuela:2017kcg,Croon:2018ftb,Bezares:2018qwa,Sanchis-Gual:2018oui}. Moreover, since bosonic stars can achieve a compactness comparable to that of black holes, they provide a case study example of black hole mimickers, $e.g.$~\cite{Eilers:2013lla,Vincent:2015xta,Cao:2016zbh,Shen:2016acv,Grould:2017rzz,Olivares:2018abq}. Finally, in the last few years there has been considerable interest in a possible role of boson stars in the dark matter problem, in particular in connection to ultralight bosonic dark matter candidates~\cite{Suarez:2013iw,Hui:2016ltb}.

Amongst the theoretically suggested ultralight dark matter particles, axion like particles (ALPs) are alongside the best motivated ones. 
The Quantum Chromo Dynamics (QCD) axion \cite{Weinberg:1977ma,Wilczek:1977pj} is a pseudo Nambu-Goldstone boson suggested by the Peccei-Quinn mechanism \cite{Peccei:1977hh} proposed to solve the strong CP problem in QCD \cite{tHooft:1976snw,Jackiw:1976pf,Callan:1979bg}. 
More recently, the axion started to be used as a prototype for weakly-interacting ultralight bosons beyond Standard Model \cite{Jaeckel:2010ni,Essig:2013lka,Goodsell:2009xc,Arvanitaki:2010sy} and considered as a plausible dark matter candidate. In particular, ALPs are ubiquitous in string compactifications, wherein they can be seen both as particles beyond Standard Model and signature of extra dimensions \cite{Svrcek:2006yi, Arvanitaki:2009fg}. 

Recently, spherical boson stars with self-interactions determined by an axion-type potential were discussed in~\cite{Guerra:2019srj} and dubbed \textit{axion boson stars}.\footnote{Axionic stars with a real scalar field (oscillatons) had been previously considered, $e.g.$~\cite{Hertzberg:2018lmt,Visinelli:2017ooc}.}  The goal of this paper is to construct and study the basic physical properties of \textit{rotating axion boson stars} (RABSs), the spinning generalisation of the solutions in \cite{Guerra:2019srj}. On the one hand, astrophysical objects in the Universe generically have angular momentum; thus, it is important to consider RABSs to assess the physical plausibility of axion bosonic stars. On the other hand, numerical evolutions have recently revealed an instability of the simplest model of spinning boson stars~\cite{Sanchis-Gual:2019ljs}. This instability could be mitigated by self-interactions, thus adding another motivation to construct RABS.

This paper is organised as follows: In Section 2, we present the full model, the equations of motion, the \textit{ansatz} used to solve them, and the QCD axion potential. In Section 3, we provide the numerical framework in which we base our results, discuss the boundary conditions, the quantities of interest and the numerical approach. In Section 4, we present the numerical results, describing the space of  RABSs solutions and analysing some physical properties. We conclude our paper in Section 5 with some remarks.

\section{The model}

We will work within the Einstein-Klein-Gordon theory, which describes a massive complex scalar field, $\Psi$, minimally coupled to Einstein's gravity. The action of the model is, using units such that $G = c = \hbar = 1$,
\begin{equation}\label{Eq:Action}
	\mathcal{S} = \int d^4 x \sqrt{-g} \left[ \frac{R}{16\pi} - g^{\mu\nu} \partial_{\mu} \Psi^* \partial_{\nu} \Psi - V(|\Psi|^2) \right] \ ,
\end{equation}
where $R$ is the Ricci scalar, $\Psi$ is the complex scalar field and $V$ is the scalar self-interaction potential. The equations of motion resulting from the variation of the action with respect to the metric and scalar field, are,
\begin{eqnarray}
	&E_{\mu\nu} \equiv R_{\mu\nu} - \frac{1}{2} g_{\mu\nu} R - 8\pi \ T_{\mu\nu} = 0 \ ,& \label{Eq:FieldEquationsMetric} \\
	&\Box \Psi - \dfrac{\partial V}{\partial |\Psi|^2} \Psi = 0 \ ,& \label{Eq:FieldEquationsScalarField}
\end{eqnarray}
where 
\be
T_{\mu\nu} = 2 \partial_{(\mu} \Psi^* \partial_{\nu)} \Psi - g_{\mu\nu} \left( \partial^\alpha \Psi^* \partial_\alpha \Psi + V\right) \ ,
\ee
is the energy-momentum tensor associated with the scalar field.

A global $U(1)$ transformation $\Psi \rightarrow e^{i\alpha} \Psi$, where $\alpha$ is a constant, leaves the above action invariant; thus it is possible to write a scalar 4-current \cite{Herdeiro:2014goa}, $j^\mu = -i \left( \Psi^* \partial^\mu \Psi - \Psi \partial^\mu \Psi^* \right)$ which is conserved $D_\mu j^\mu = 0$. The existence of this symmetry and conserved current implies the existence of a conserved quantity -- the \textit{Noether charge} -- that can be computed by integrating the timelike component of the 4-current,
\begin{equation}
	Q = \int_\Sigma j^t \ .
\end{equation}
This quantity can be interpreted as the number of scalar particles in a given solution, albeit this relation only becomes rigorous upon field quantisation.  A simple computation, moreover, shows that for solitonic solutions (without an event horizon) $Q$ is related with the total angular momentum as~\cite{Yoshida:1997qf,Schunck:1996he} 
\be
J = m Q \ .
\ee
This is a generic relation for rotating boson stars, already observed in other models with a self-interactions potential - see, $e.g.$~\cite{Kleihaus:2007vk}.

We are interested in horizonless, everywhere regular, axisymmetric and asymptotically flat solutions of the above field equations. To do so, we choose an \textit{ansatz} for the metric and scalar field typical for rotating boson star models, see $e.g.$~\cite{Herdeiro:2015gia,Herdeiro:2015tia}:
\begin{eqnarray}
	&ds^2 = -e^{2F_0(r,\theta)} dt^2 + e^{2F_1(r,\theta)} \left( dr^2 + r^2 d\theta^2 \right) + e^{2F_2(r,\theta)} r^2 \sin^2 \theta \left( d\varphi - \dfrac{W(r,\theta)}{r} dt \right)^2  , \ \ & \label{Eq:AnsatzMetric} \\
	&\Psi = \phi(r,\theta) e^{i(\omega t - m \varphi)} \ ,& \label{Eq:AnsatzScalarField}
\end{eqnarray}
where $F_i, W$ and $\phi$ are \textit{ansatz} functions that only depend on ($r,\theta$) coordinates; $\omega$ and $m = \pm 1, \pm 2, \dots$ are the angular frequency and azimuthal harmonic index of the scalar field, respectively.

To obtain solutions one has to specify the scalar field potential $V$. Following~\cite{Guerra:2019srj}, in this work we shall use the QCD axion potential - see~\cite{diCortona:2015ldu} for a discussion -, added to a constant term, in order to construct asymptotically flat RABS. The potential to be considered is
\begin{equation}
\label{Eq:Potential}
	V(\phi) = \frac{2 \mu_a^2 f_a^2}{B} \left[ 1 - \sqrt{1 - 4 B \sin^2 \left( \frac{\phi}{2 f_a} \right)} \right] \ ,
\end{equation}
where $B$ is a constant, defined in terms of the up and down quarks' masses, $m_u,m_d$, as $B = \frac{z}{(1+z)^2} \approx 0.22$, where $z \equiv m_u/m_d \approx 0.48$. The potential has two free parameters, $\mu_a$ and $f_a$. To see their meaning, we expand the potential around $\phi = 0$, yielding
\begin{equation}\label{Eq:PotentialApprox}
	V(\phi) = \mu_a^2 \phi^2 - \left( \frac{3B-1}{12} \right) \frac{ \mu_a^2}{f_a^2} \phi^4 + \dots
\end{equation}
where one can identify the ALP has mass and quartic self-interaction coupling, respectively,
\be
m_a =  \mu_a \ , \qquad \lambda_a = -\left( \frac{3B-1}{12} \right) \frac{ \mu_a^2}{f_a^2} \ .
\ee 
Thus,  $\mu_a$ defines the mass of the ALP and $f_a$ defines the self-interactions.
We shall refer to $\mu_a$ and $f_a$ as the ALP mass and decay constant, respectively.
We remark the above expansion is only valid if 
\be
f_a \gg \phi \ .
\label{miniap}
\ee
In fact, to leading order, only the mass term remains. Thus, we can anticipate that the properties of the solutions will match those of the standard spinning mini-boson stars~\cite{Yoshida:1997qf,Schunck:1996he}  when $f_a$ is large. 

\begin{figure}[ht!]
\begin{center}
\includegraphics[height=.280\textheight, angle =0]{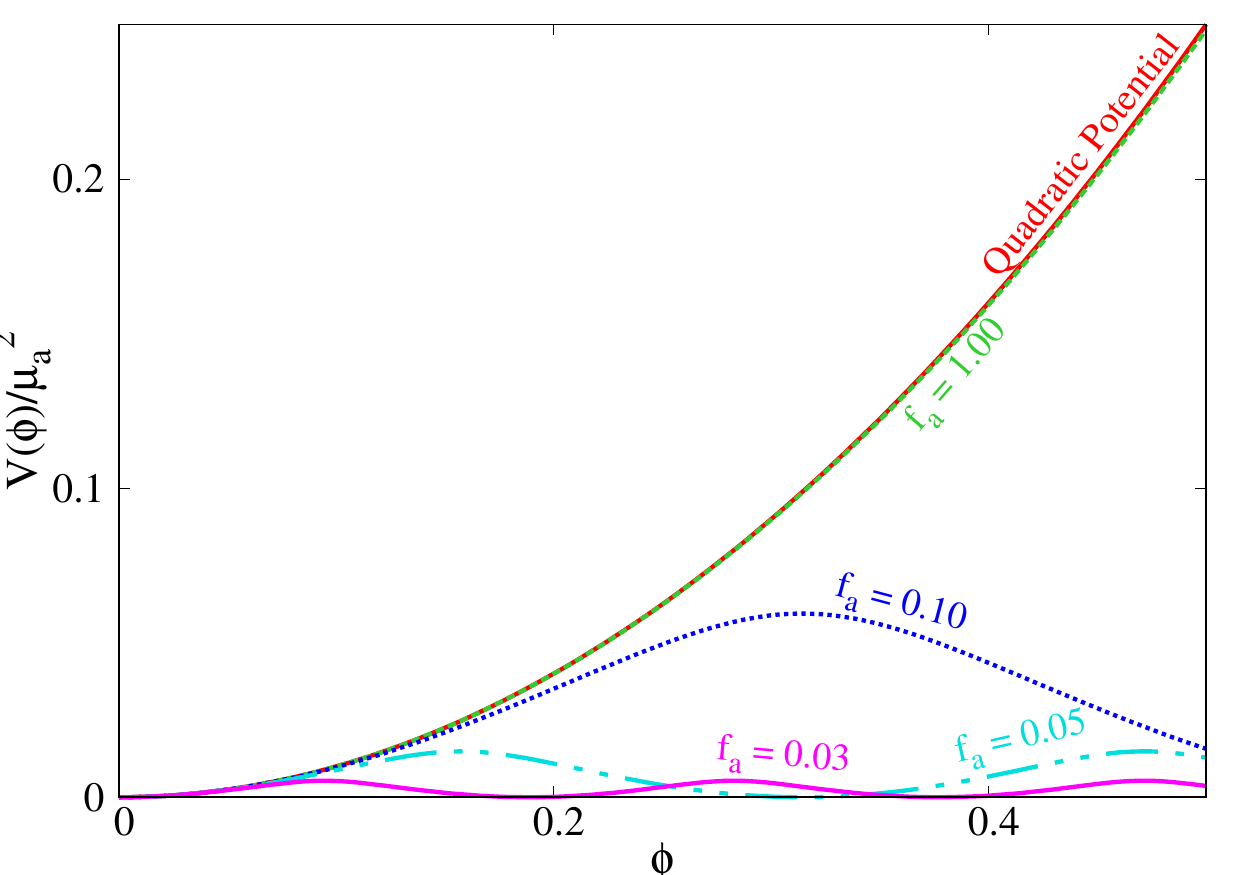}
\end{center}
  \vspace{-0.5cm}
\caption{Axion potential~\eqref{Eq:Potential} for several values of the decay constant $f_a$ together with the quadratic potential.}
\label{Fig:PotentialProfile}
\end{figure}

In Fig.~\ref{Fig:PotentialProfile} we show the axion potential~\eqref{Eq:Potential} for several values of $f_a$, together with the quadratic potential - first term in \eqref{Eq:PotentialApprox}. It can be seen that for $f_a=1.0$ the quadratic potential is a good approximation for the range of values of $\phi$ displayed, which will be in the range obtained in the RABSs solutions below, $cf.$  top panel in Fig. \ref{Fig:MaxPhi} below.

We remark that the potential 
(\ref{Eq:Potential})
satisfies the conditions in Ref.
\cite{Coleman:1985ki}
allowing for the existence of solitonic solutions
even in the absence of the gravity term in the action (\ref{Eq:Action}).
Our results indicate that indeed, 
such configurations exist in a flat spacetime background,
  sharing all the basic properties of the (non-gravitating) Q-balls
with a sextic potential~\cite{Kleihaus:2005me}. 
A discussion of these solutions will be reported elsewhere.

\section{Framework}

\subsection{Boundary conditions}

With the metric and scalar field \textit{ansatz} defined in Eqs. \eqref{Eq:AnsatzMetric} and \eqref{Eq:AnsatzScalarField}, we now need to specify the boundary conditions for the \textit{ansatz} functions that best suit the problem at hand. For asymptotically flat solutions, the asymptotic behaviour ($r \rightarrow \infty$) of the functions must be,
\begin{equation}
	\lim_{r \rightarrow \infty} F_i = \lim_{r \rightarrow \infty} W = \lim_{r \rightarrow \infty} \phi = 0 \ .
\end{equation}
Axial symmetry, together with regularity on the symmetry axis, impose on the axis ($\theta = 0, \pi$),
\begin{equation}
	\partial_\theta F_i = \partial_\theta W = \partial_\theta \phi = 0 \ .
\end{equation}
Furthermore, absence of conical singularities implies also that, on the axis,
\begin{equation}
	F_1 = F_2 \ .
\end{equation}
In this work we shall focus on solutions that are symmetric \textit{w.r.t} a reflection along the equatorial plane. These are \textit{even parity solutions}. Typically this is the case for fundamental solutions, but there can be odd parity excited states. This choice means that one only has to consider the range $0 \leqslant \theta \leqslant \pi/2$. This also implies that, on the equatorial plane, $\theta = \pi/2$,
\begin{equation}
	\partial_\theta F_i = \partial_\theta W = \partial_\theta \phi = 0  \ .
\end{equation}
At the origin, $r \rightarrow 0$, we impose the following boundary conditions,
\begin{equation}
	\partial_r F_i = W = \phi = 0 \ ,
\end{equation}
which are dictated by regularity therein.

\subsection{Extracting mass, angular momentum, radius and compactness}

For RABSs, two of the key quantities of interest are encoded in the decay of the metric functions towards spatial infinity. Such quantities are the ADM mass $M$ and angular momentum $J$ which are read off from the asymptotic  behaviour of the following metric functions,
\begin{equation}
	g_{tt} = -e^{2F_0} + e^{2F_2} W^2 \sin^2 \theta = -1 + \frac{2M}{r} + \dots \hspace{5pt}, \hspace{5pt} g_{\varphi t} = -e^{2F_2} W r \sin^2 \theta = - \frac{2J}{r} \sin^2 \theta + \dots \ .
\end{equation}
Alternatively, both quantities can be computed through their Komar integrals \cite{Poisson:2009pwt},
\begin{equation}
	M = - 2 \int_\Sigma d S_\mu \left( T^\mu_\nu t^\nu - \frac{1}{2} T t^\mu \right) \hspace{5pt} , \hspace{5pt} J = \int_\Sigma d S_\mu \left( T^\mu_\nu \varphi^\nu - \frac{1}{2} T \varphi^\mu \right) \ ,
\end{equation}
where $\Sigma$ is a spacelike surface bounded by a 2-sphere at infinity, $S_\infty^2$, and $t^\mu$ and $\varphi^\mu$ are the timelike and rotational Killing vectors, respectively. A good test for the numerical results amounts to checking the match between the mass and angular momentum obtained from the asymptotic behaviour and from the Komar integrals.

Another quantity of interest is the compactness of the RABSs. Since the scalar field decays exponentially, these stars do not have a well defined surface, wherein a discontinuity in the matter distribution occurs. Nonetheless, a standard definition of "surface" in this context is as follows \cite{Herdeiro:2015gia}. Firstly, we compute the perimeteral radius $R_{99}$, which contains 99\% of the total mass of the BS, $M_{99}$. The perimetral radius is a geometrically significant radial coordinate $R$, such that a circumference along the equatorial plane has perimeter $2\pi R$, and it is related to the radial coordinate $r$ in the \textit{ansatz} metric by $R = e^{F2}r$.  Secondly, we define the inverse compactness as the ratio between $R_{99}$ and the Schwarzschild radius associated with 99\% of the RABS's mass, $R_{Schw} = 2M_{99}$,
\begin{equation}
	\text{Compactness}^{-1} = \frac{R_{99}}{2 M_{99}} \ .
\end{equation}
We expect stars to be less compact than black holes and thus this quantity to be larger than unity for RABSs.

\subsection{Numerical approach}

In order to perform the numerical integration of the equations resulting from \eqref{Eq:FieldEquationsMetric}-\eqref{Eq:FieldEquationsScalarField} with the \textit{ansatz} Eqs. \eqref{Eq:AnsatzMetric} and \eqref{Eq:AnsatzScalarField}, it is useful to rescale key quantities by $\mu_a$,
\begin{equation}
	r \rightarrow r \mu_a \hspace{10pt} , \hspace{10pt} \phi \rightarrow \phi\sqrt{4\pi} \hspace{10pt}, \hspace{10pt} \omega \rightarrow \omega/\mu_a \hspace{10pt} \ .
\end{equation}
As a result, the dependency on $\mu_a$ disappears from the equations, and the global quantities will be express in "units" of $\mu_a$.

In our approach, the field equations reduce to a set of five coupled, non-linear, elliptic partial differential equations for the functions $\mathcal{F}_a = (F_0, F_1, F_2, W, \phi)$. They consist of the Klein-Gordon equation \eqref{Eq:FieldEquationsScalarField} together with suitable combinations of the Einstein equations \eqref{Eq:FieldEquationsMetric},
\begin{eqnarray}
	&&E^r_r + E^\theta_\theta - E^\varphi_\varphi - E^t_t = 0 \ , \\
	&&E^r_r + E^\theta_\theta - E^\varphi_\varphi + E^t_t + 2 W E^t_\varphi = 0 \ , \\
	&&E^r_r + E^\theta_\theta + E^\varphi_\varphi - E^t_t - 2 W E^t_\varphi = 0 \ , \\
	&&E^t_\varphi = 0 \ .
\end{eqnarray}
The remaining two equations $E^r_\theta = 0$ and $E^r_r - E^\theta_\theta = 0$ are not solved directly; instead, they are used as constraints to monitor the numerical solution. Typically, these constraints are satisfied at the level of the overall numerical accuracy. We shall not exhibit the explicit equations here. They are the same, up to the potential terms, as the ones in Section 2.3 of~\cite{Herdeiro:2015gia}.

Our numerical treatment is as follows. The first step is to define a new radial coordinate that can map the semi-infinite region $[0,\infty)$ to the finite region $[0,1]$. Such coordinate can be defined as $x \equiv r/(1+r)$. The second step is to discretise the equations for $\mathcal{F}_a$ on a grid in $x$ and $\theta$. For the majority of the results in this work, we used an equidistant grid with 401$\times$40 points. The grid covers the integration region $0 \leqslant x \leqslant 1$ and $0 \leqslant \theta \leqslant \pi/2$.

The equations for $\mathcal{F}_a$ have been solved subject to the boundary conditions mentioned before. All numerical calculations have been performed by using a professional package entitled \textsc{fidisol/cadsol}~\cite{schoen}, which uses a Newton-Raphson method with an arbitrary grid and arbitrary consistency order. This package also provides an error estimate for each unknown function. For the RABSs solutions in this work, the maximal numerical error is estimated to be on the order of $10^{-3}$. As mentioned in the previous subsection, a further test of the numerical accuracy of the solutions can be obtained by comparing the ADM mass and angular momentum with the corresponding Komar integral. This comparison yields an error estimate of the same order as the one given by the solver. 

In our scheme, there are three input parameters: \textbf{i}) the decay constant $f_a$ in the potential \eqref{Eq:Potential}; \textbf{ii}) the angular frequency of the scalar field $\omega$ and \textbf{iii}) the azimuthal harmonic index $m$. The number of nodes $n$ of the scalar field, as well as all other quantities of interest mentioned before, are computed from the numerical solution. For simplicity, we have restricted our study to fundamental configurations, \textit{i.e.} with a nodeless scalar field, $n=0$ and with $m=1$.

\section{Numerical Results}

\begin{figure}[ht!]
\begin{center}
\includegraphics[height=.230\textheight, angle =0]{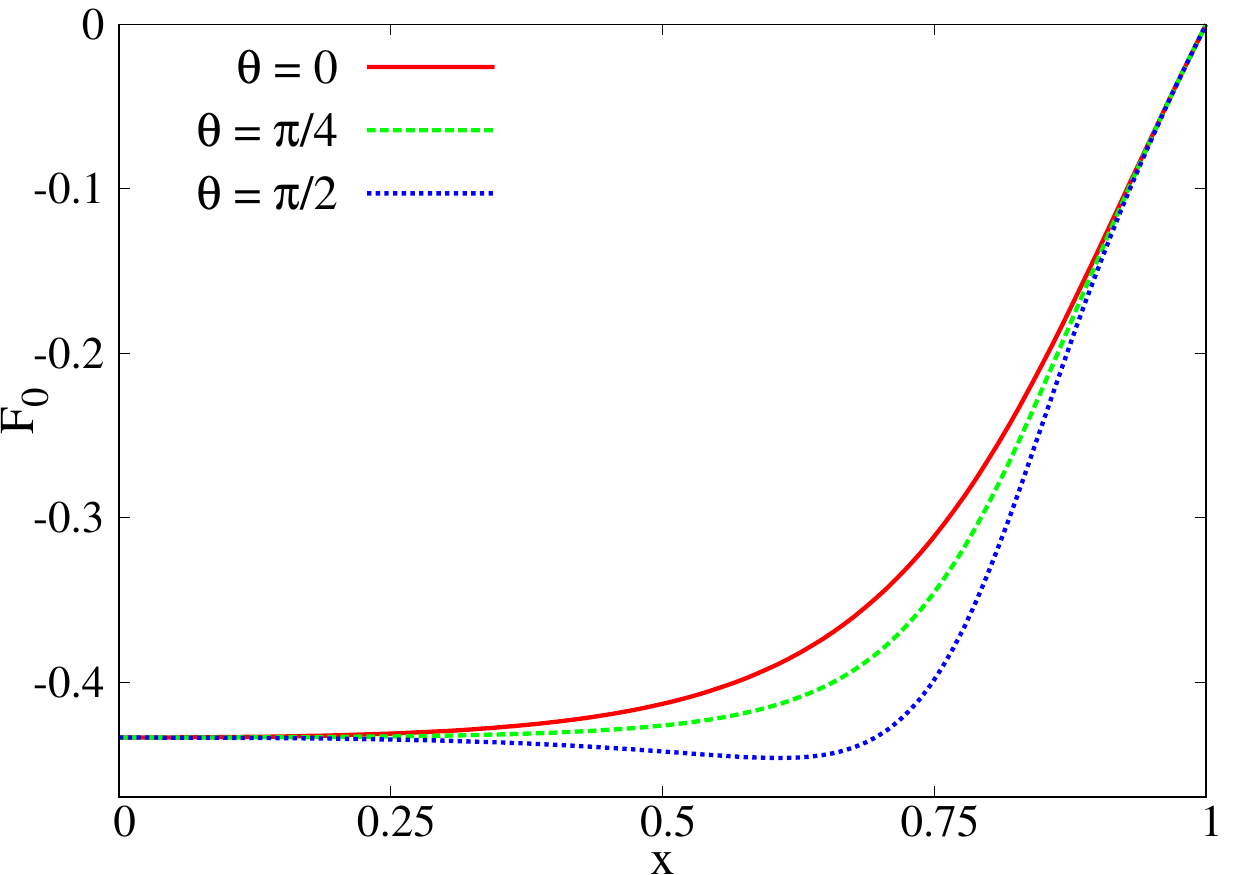} 
\includegraphics[height=.230\textheight, angle =0]{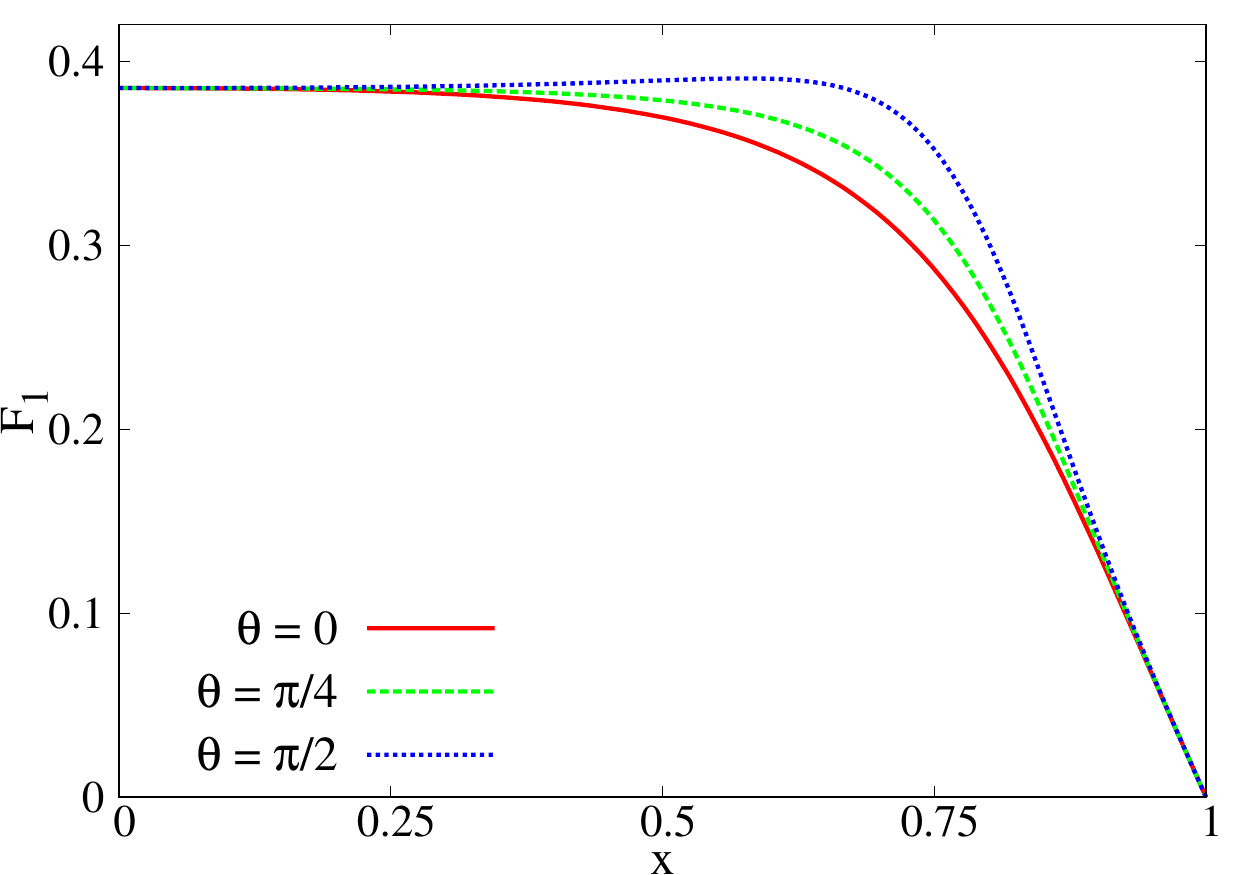} \ \ 
\includegraphics[height=.230\textheight, angle =0]{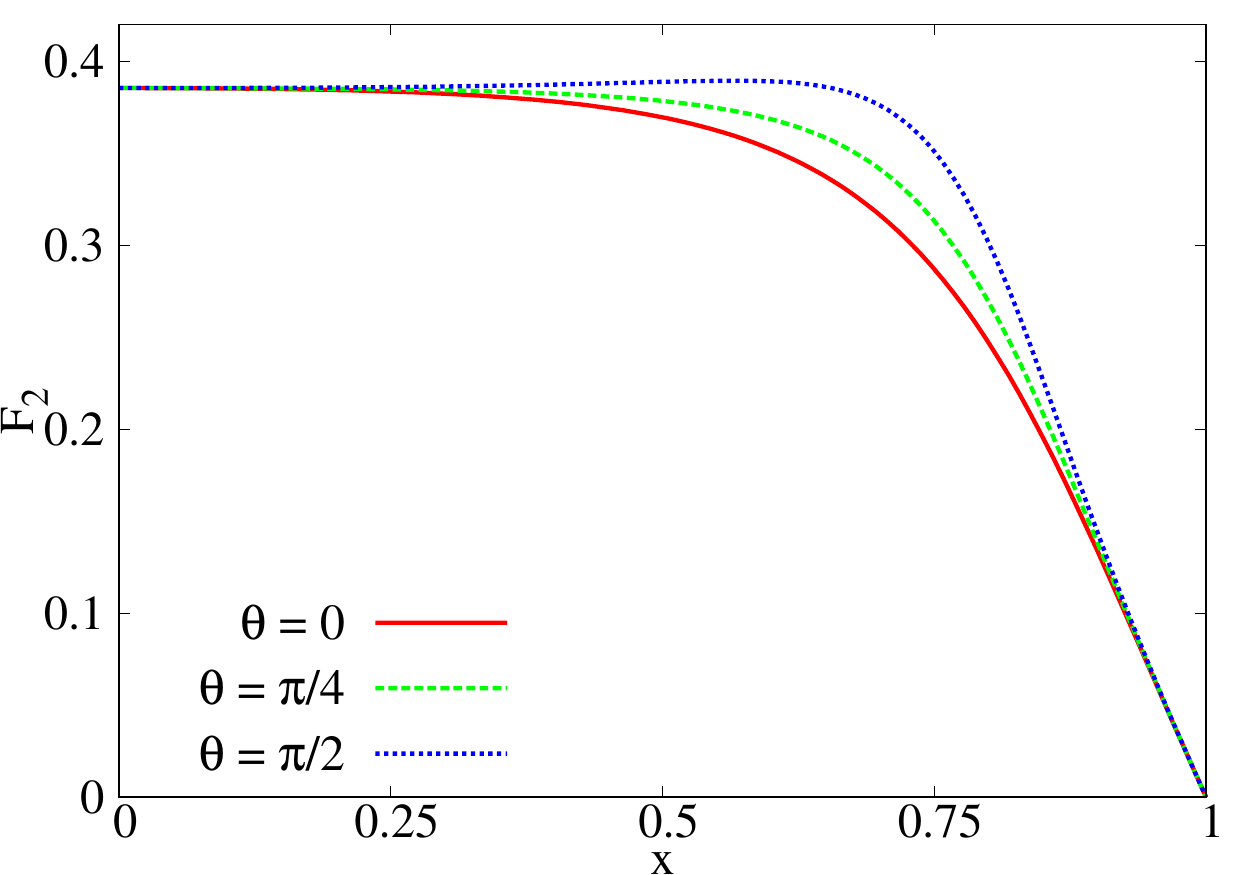} 
\includegraphics[height=.230\textheight, angle =0]{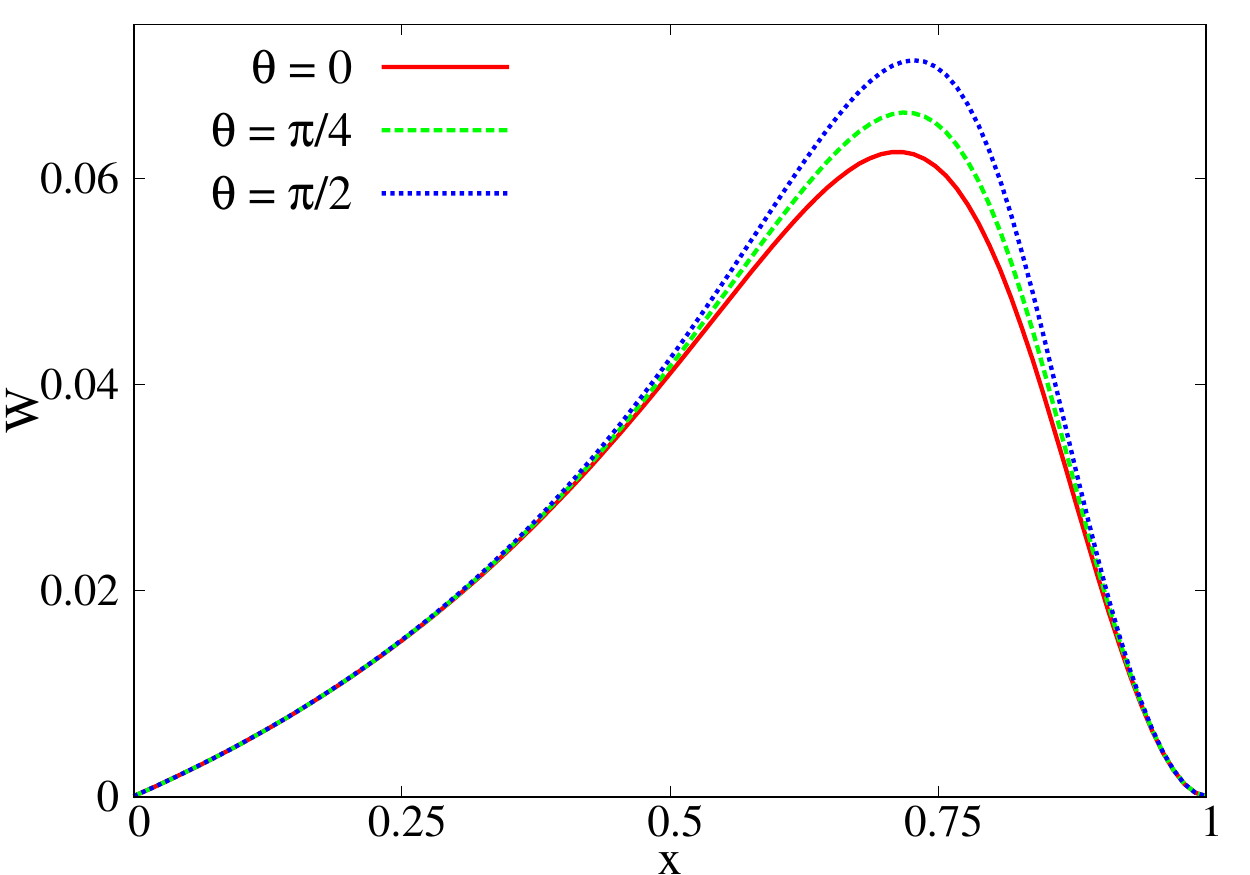} \ \
\includegraphics[height=.230\textheight, angle =0]{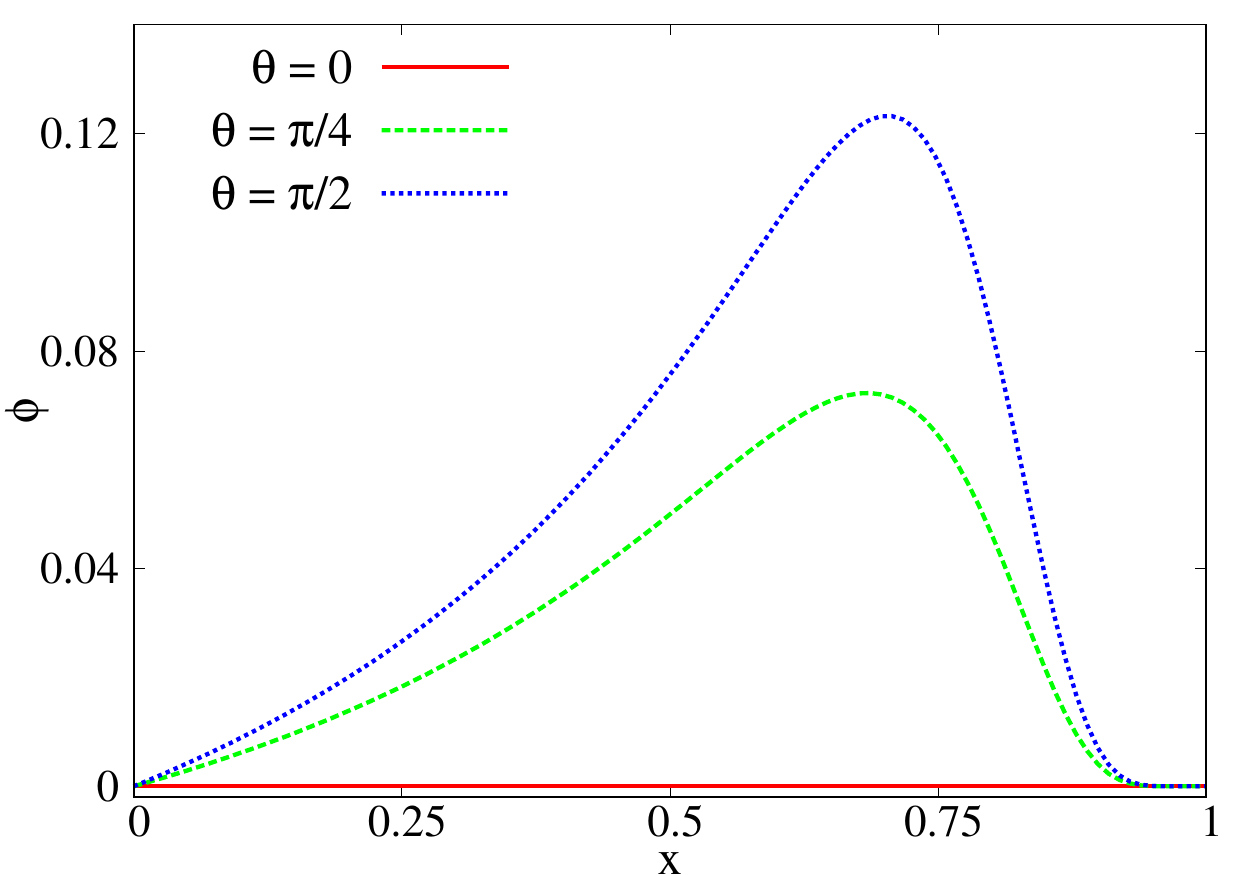} 
\includegraphics[height=.230\textheight, angle =0]{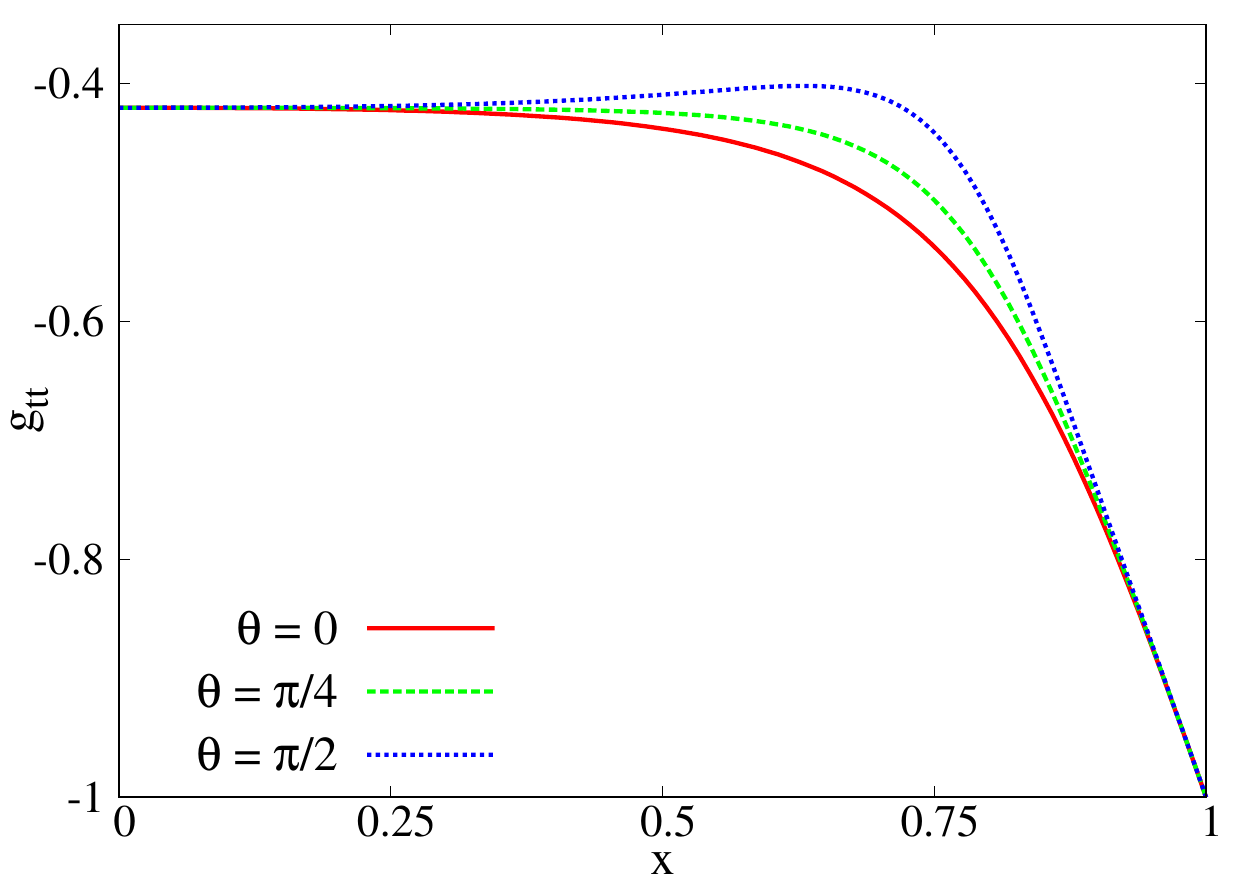} \ \
\end{center}
  \vspace{-0.5cm}
\caption{Profile functions of a typical solution 
with $f_a=1.00$ and $\omega=0.80$,
$vs.$ $x = r/(1+r)$, which compactifies the exterior region,
 for three different polar angles $\theta$.
}
\label{Fig:MetricFunctions}
\end{figure}

\begin{figure}[ht!]
\begin{center}
\includegraphics[height=.230\textheight, angle =0]{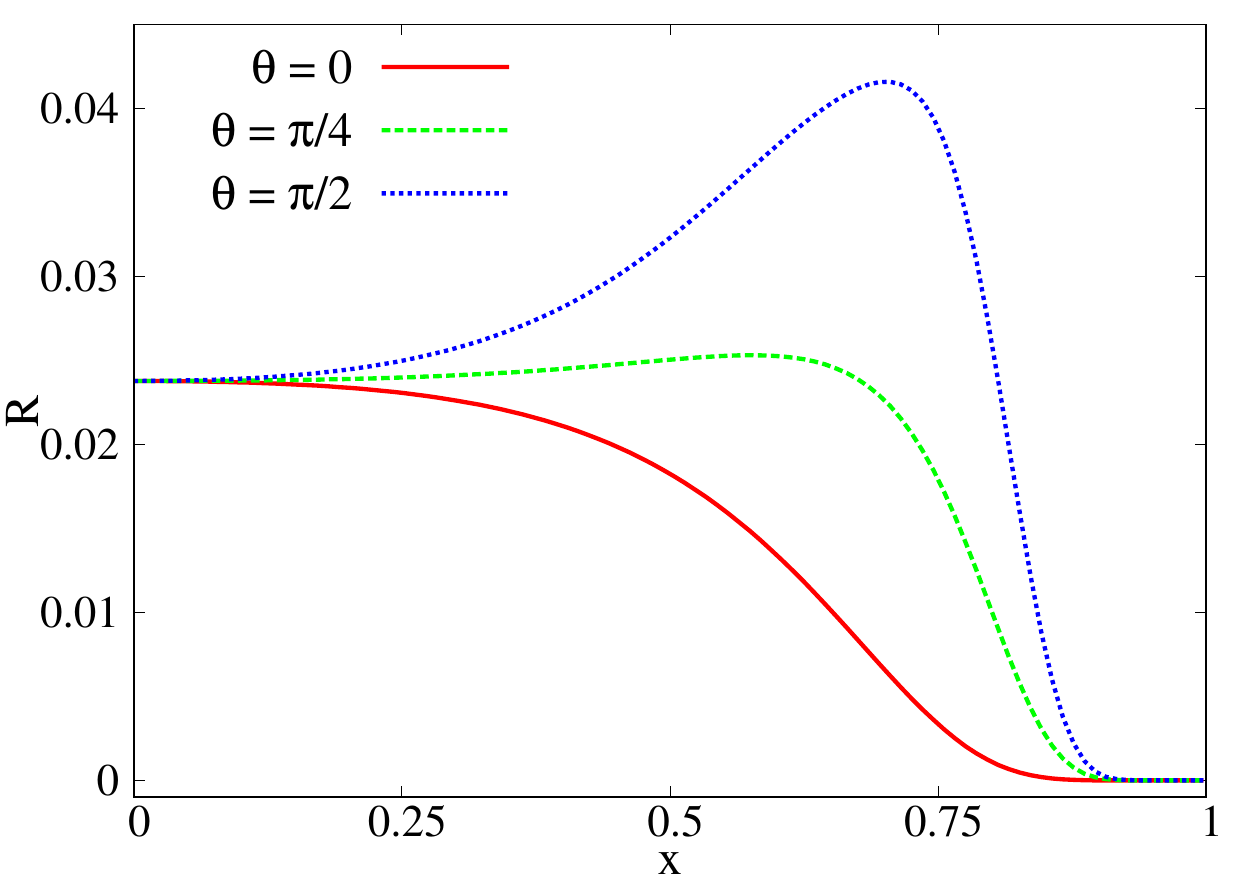} 
\includegraphics[height=.230\textheight, angle =0]{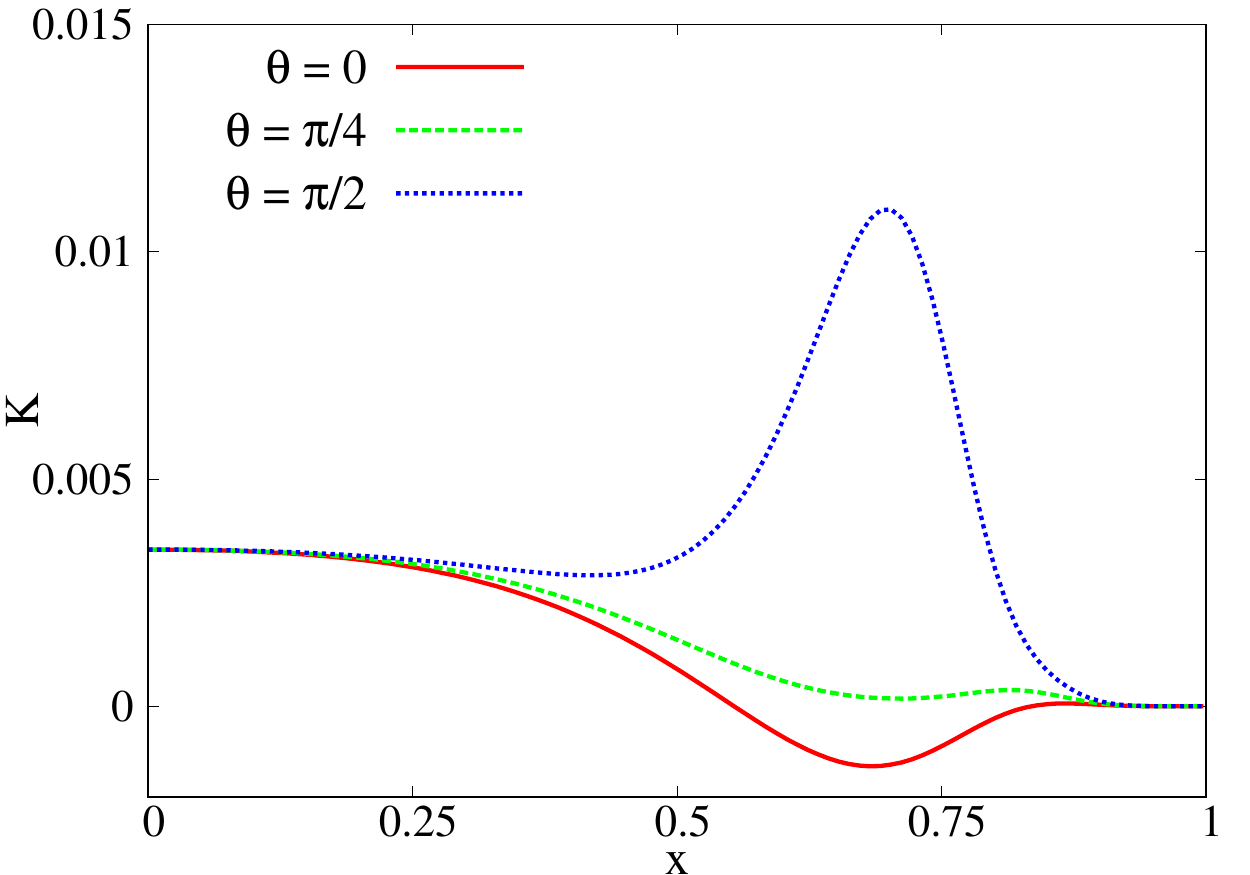} \ \ 
\includegraphics[height=.230\textheight, angle =0]{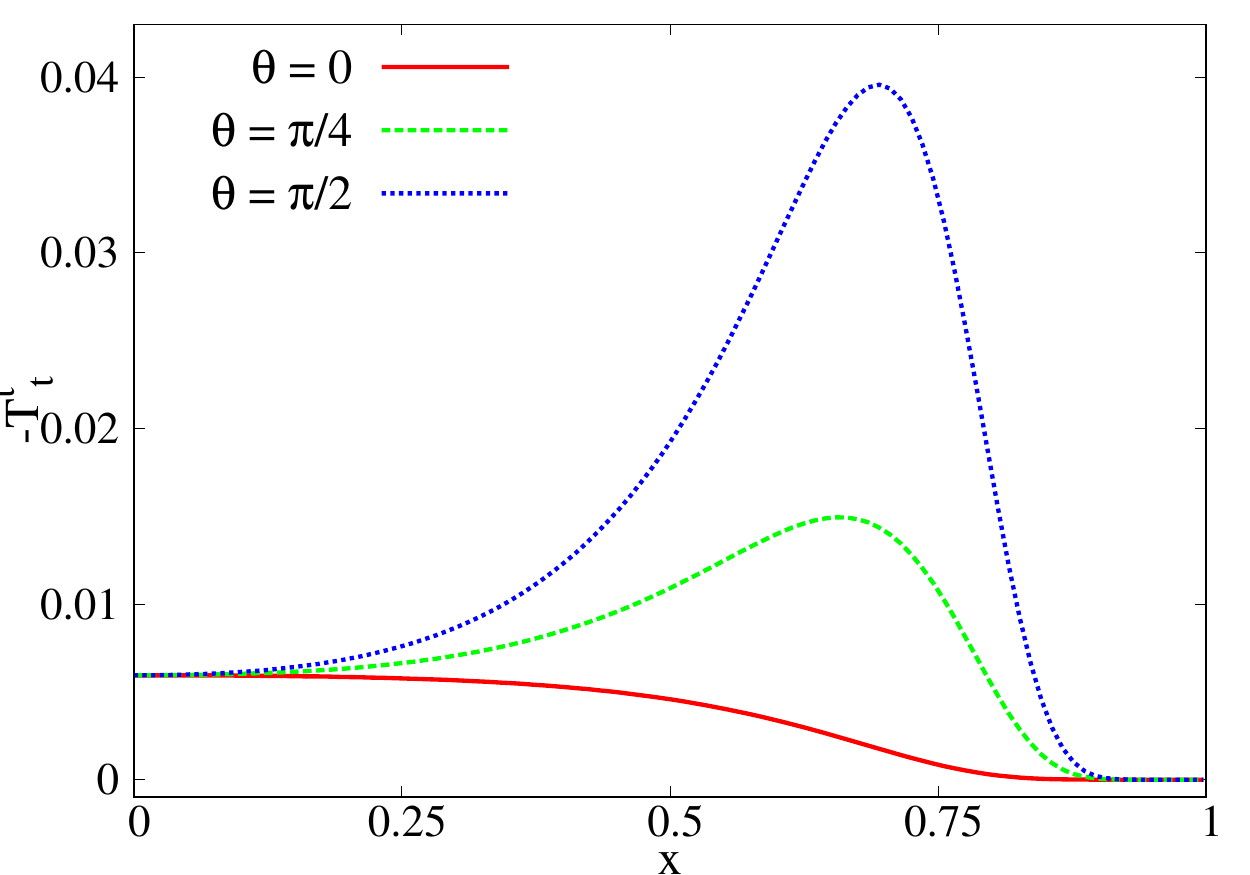} 
\includegraphics[height=.230\textheight, angle =0]{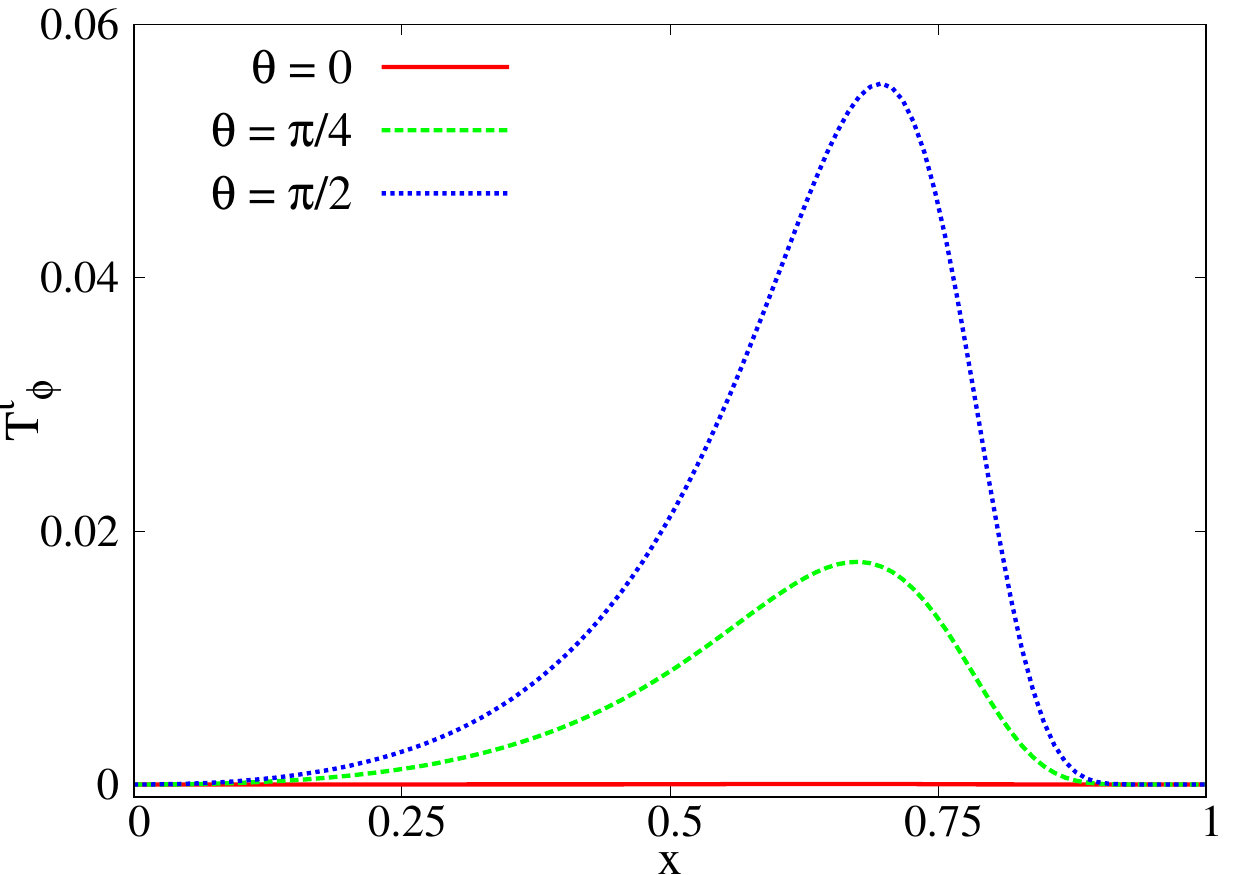} \ \
\end{center}
  \vspace{-0.5cm}
\caption{The Ricci $R$ and Kretschmann $K$ scalars and the components $T^t_t$ and $T^t_\varphi$ of the energy-momentum tensor
$vs.$ $x \equiv r/(1+r)$, of a typical solution 
with $f_a=1.00$ and $\omega=0.80$
 for three different polar angles $\theta$.
}
\label{Fig:EnergyGeometricFunctions}
\end{figure}

We obtained around sixty thousand solutions, considering a set of illustrative values of the decay constant $f_a$. For each choice of $f_a$, $\omega$ spans an interval of values. For all solutions, the metric functions $\mathcal{F}_a$, together with their first and second derivatives with respect to both $r$ and $\theta$, have smooth profiles. This leads to finite curvature invariants on the full domain of integration. The profile of the metric functions, together with the Ricci and Kretschmann scalars, $R$ and $K \equiv R_{\alpha\beta\mu\nu}R^{\alpha\beta\mu\nu}$, as well as the components $T_t^t$ and $T_\varphi^t$ of energy-momentum tensor, of a typical solution are exhibited in Figs.~\ref{Fig:MetricFunctions} and \ref{Fig:EnergyGeometricFunctions}. In particular, observe the toroidal shape of the scalar field (Fig. \ref{Fig:MetricFunctions}, bottom left panel). Thus, similarly to mini-BSs, RABSs are mass tori in General Relativity. There is a clear imprint of this toroidal distribution in the curvature and energy-momentum tensor, as can be appreciated from Fig.~ \ref{Fig:EnergyGeometricFunctions}.

\subsection{The domain of existence}

In scanning the space of solutions of RABSs, after fixing the number of nodes $n=0$ and the azimuthal harmonic index $m=1$, such as to look for fundamental states, we have to vary the axion decay constant $f_a$. For each $f_a$, the domain of existence is obtained by varying the angular frequency of the scalar field $\omega$. When the decay constant is large, $f_a \rightarrow \infty$, the model reduces to that of a massive, free complex scalar field -- \textit{cf.} Eq. \eqref{Eq:PotentialApprox}, and the solutions match the usual spinning mini-boson stars. The opposite limit, wherein $f_a \rightarrow 0$, is trickier and solutions cannot be obtained for arbitrarily low $f_a$. 
For the following discussion, we have chosen a sample of illustrative values for the decay constant $f_a$ and have analysed the resulting solutions. The sample of $f_a$ values is 
\be
f_a = \{1.00, 0.10, 0.05, 0.03\} \ .
\ee

\begin{figure}[ht!]
\begin{center}
\includegraphics[height=.38\textheight, angle =0]{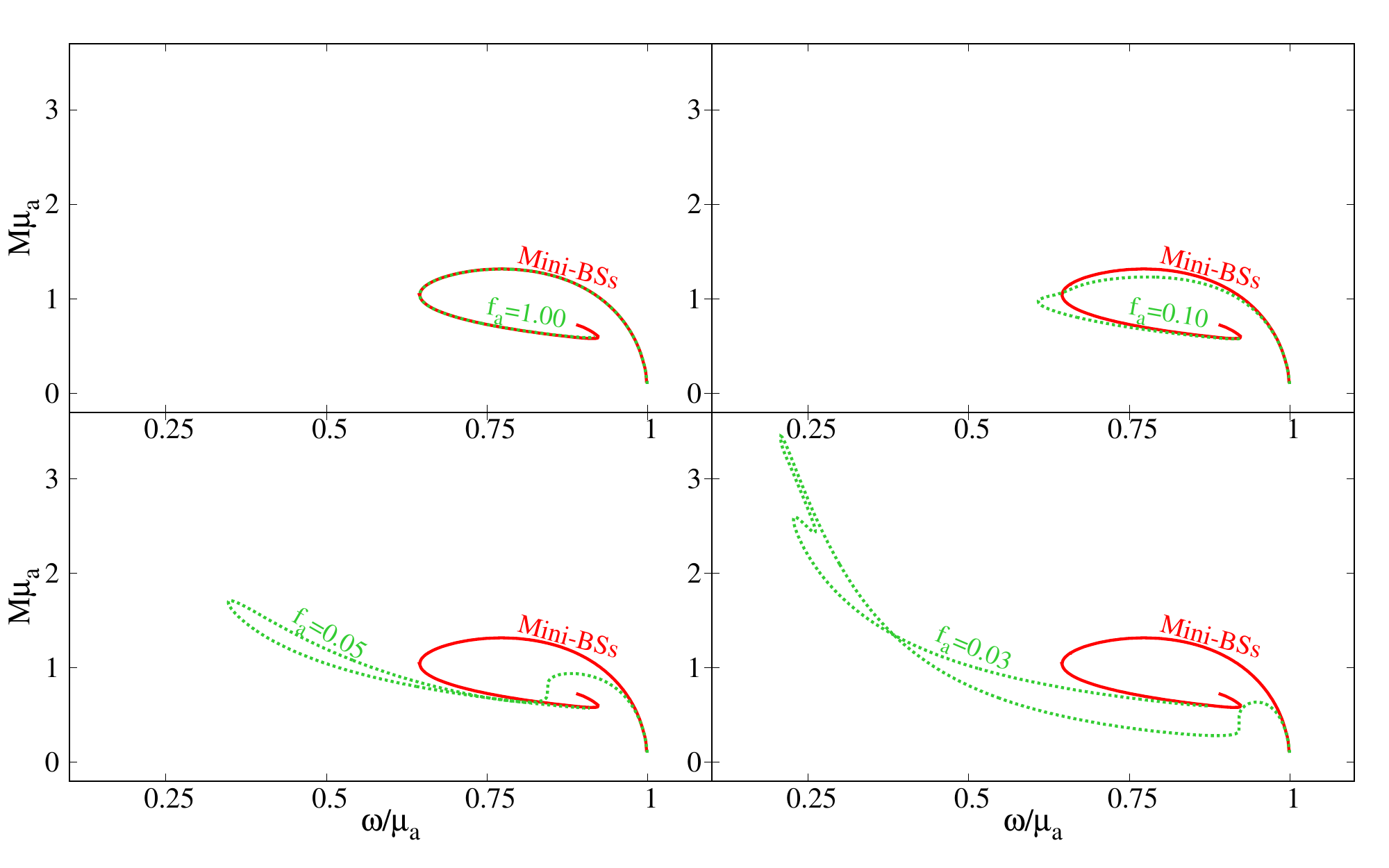} \\  \vspace{-0.5cm}
\includegraphics[height=.38\textheight, angle =0]{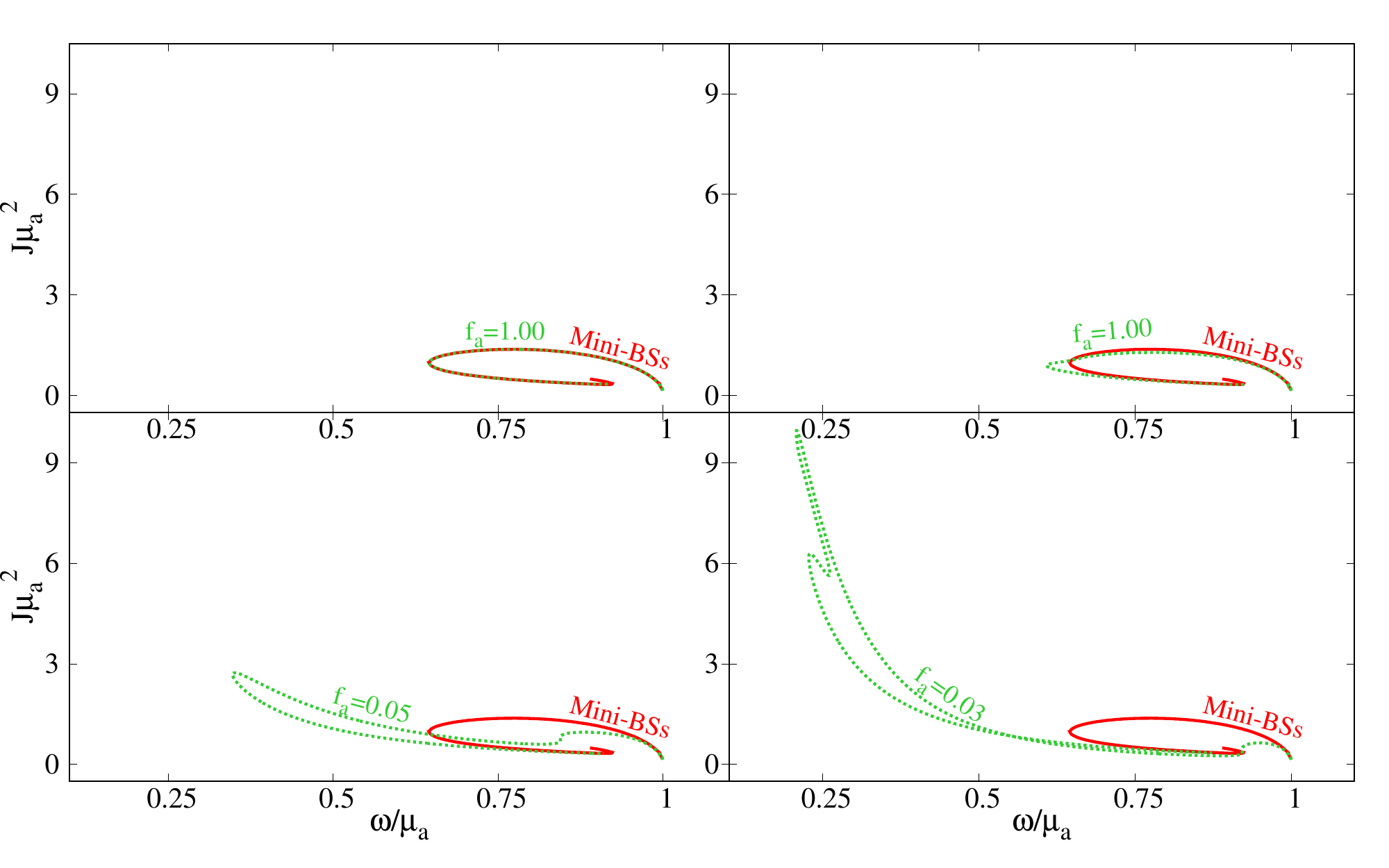}
\end{center}
  \vspace{-1cm}
\caption{ADM mass $M$ (top four panels) and total angular momentum $J$ (bottom four panels) $vs.$ the angular frequency of the scalar field $\omega$.}
\label{Fig:Mass}
\end{figure}

\begin{figure}[ht!]
\begin{center}
\includegraphics[height=.38\textheight, angle =0]{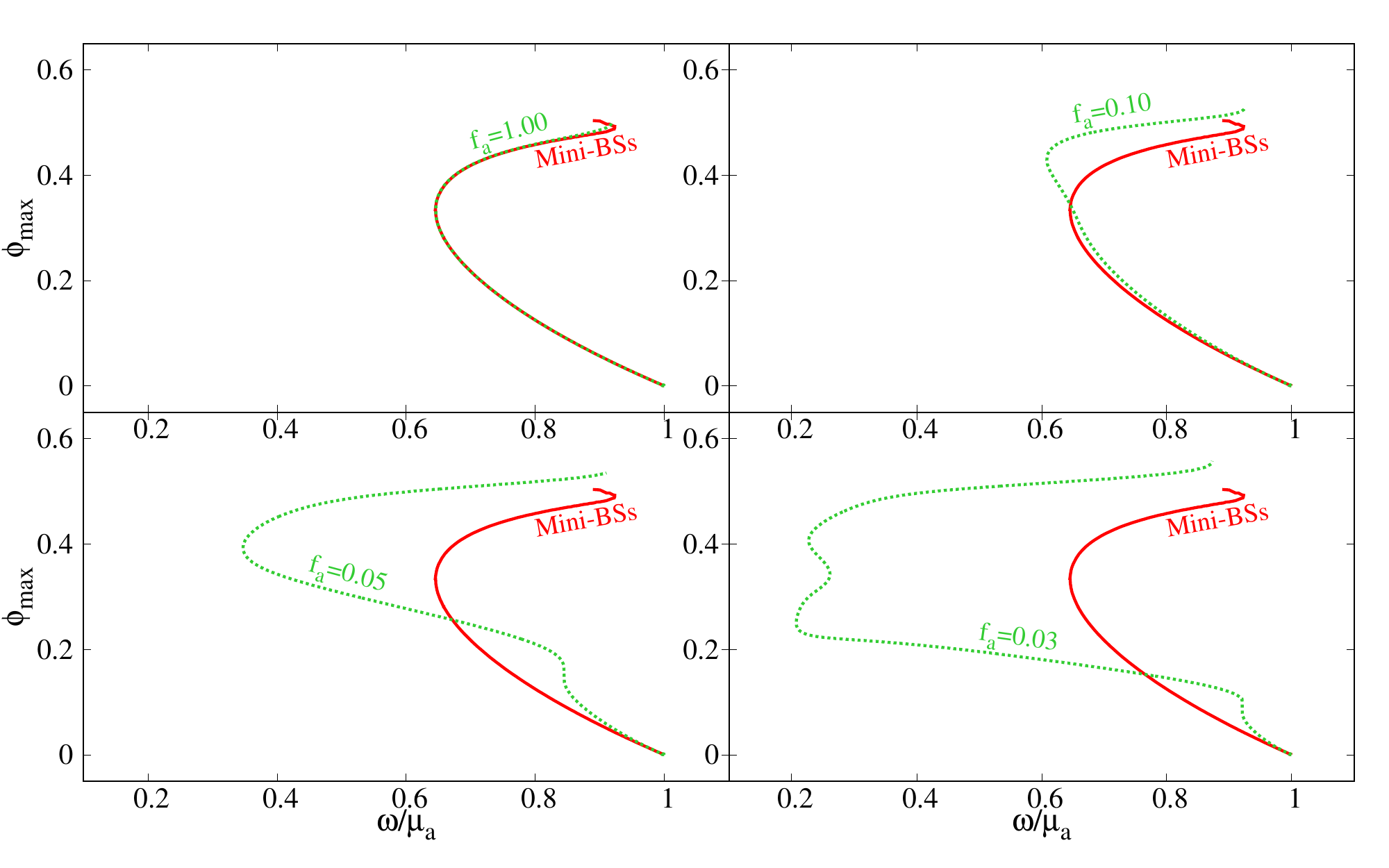} \\  \vspace{-0.5cm}
\includegraphics[height=.38\textheight, angle =0]{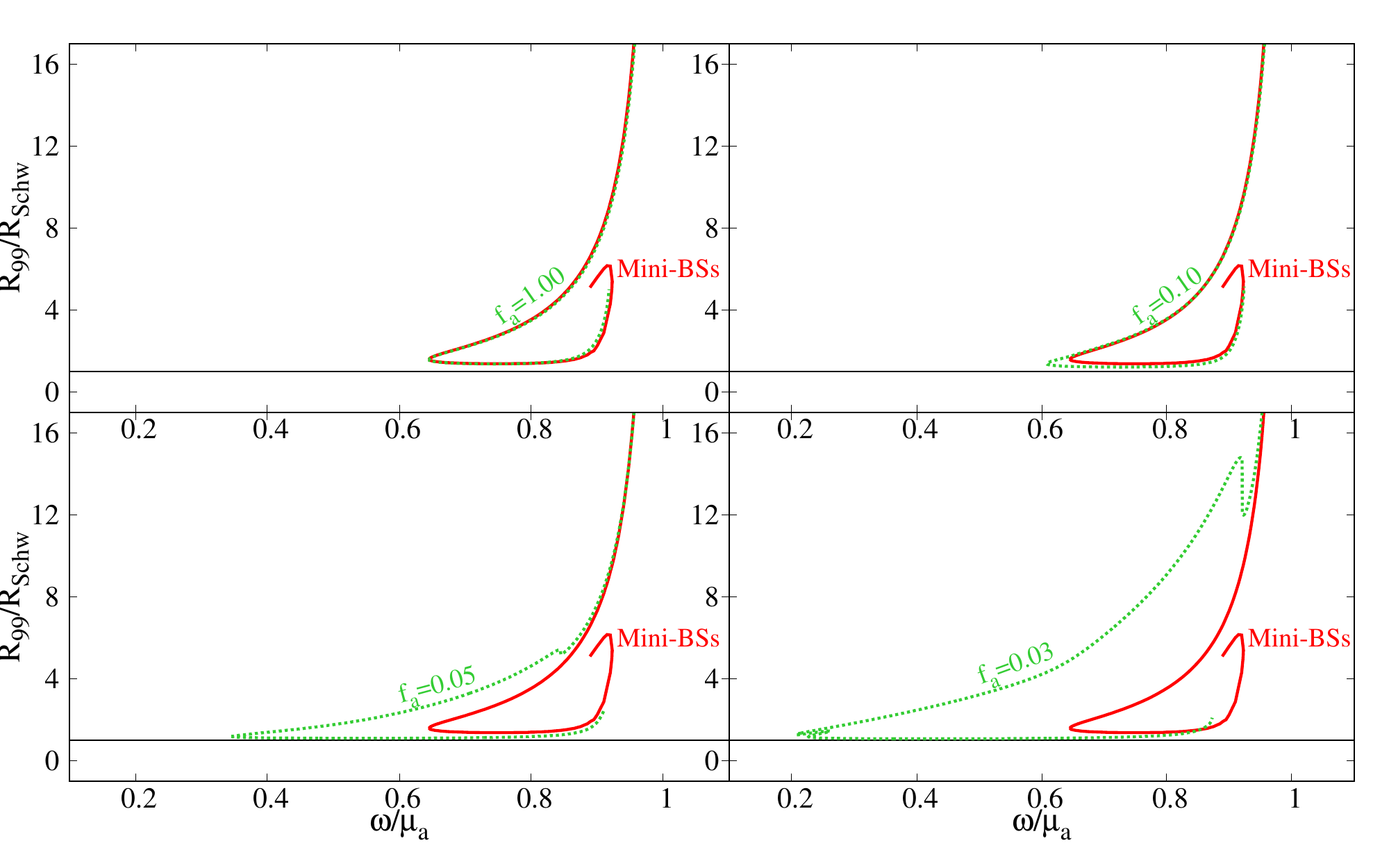}
\end{center}
  \vspace{-1cm}
\caption{Maximum value of the scalar field (top four panels) and inverse compactness (bottom four panels)  as a function of the angular frequency $\omega$. In the latter, the black horizontal lines correspond to the BH limit.}
\label{Fig:MaxPhi}

\end{figure}

In Fig.~\ref{Fig:Mass} the two global charges, mass and angular momentum, are exhibited $vs.$ the angular frequency, for the chosen sample of $f_a$ values. As a reference, all panels include as well the corresponding plot for mini-boson stars. Outstanding features include the following. Firstly, for the largest value of $f_a$ considered in the work, $f_a = 1.00$, the domain of existence of the RABSs is indistinguishable from that of mini-boson stars. This is consistent with the potential profile seen in Fig. \ref{Fig:PotentialProfile} and the range of scalar field values shown in the illustrative behaviour in Fig. \ref{Fig:MetricFunctions}, bottom left panel; one can see that $f_a = 1.00$ is large enough so that \eqref{miniap} holds. This is further confirmed by the analysis in Fig.~\ref{Fig:MaxPhi} (top panels), where the maximum value of the scalar field, $\phi_{\text{max}}$ is shown against $\omega$. Following each line starting from $\omega = 1$, $\phi_{\text{max}}$  increases monotonically, but only up to around $\phi_{\text{max}} \sim 0.5$. This holds for all numerically accurate solutions obtained. Moreover, $\phi_{\text{max}}$ is never much larger, in fact always comparable, to that in the mini-boson star case. This confirms why $f_a \gtrsim 1.00$ RABSs are essentially equivalent to mini-boson stars.

 The second feature we would like to emphasise about  Fig.~\ref{Fig:Mass} is that for all value of $f_a$, solutions exist within an interval of angular frequencies, which depends on $f_a$. As  $f_a$ decreases, the lower end of this interval decreases; then, RABSs exist with lower $\omega$ than in the mini-boson star case. This feature is instructive. In the case of quartic self-interactions, the frequency range of spinning boson stars decreases, with increasing strength of the self-interactions -- see Fig. 1 in~\cite{Herdeiro:2015tia} -- since the frequency at the lower end \textit{increases}. Such trend for the angular frequency range is never observed in the panels of Fig.~\ref{Fig:Mass}. This implies that the quartic approximation~\eqref{Eq:PotentialApprox} to the potential~\eqref{Eq:Potential} is never a good approximation for RABSs.\footnote{We remark that the sign of the quartic term in~\cite{Herdeiro:2015tia} and in~\eqref{Eq:PotentialApprox} is the opposite.}  In other words, when the quartic term becomes relevant, higher order terms become relevant as well for these solutions, in particular impacting decisively on the frequency range of the space of solutions.

A third observation concerning Fig.~\ref{Fig:Mass}, is that for small enough values of $f_a$, the line of solutions deviates from the typical spiral, as one can see for the cases of $f_a = 0.05$ and $f_a = 0.03$. The more intricate space of solutions impacts, in particular, on the global maximum of the mass. For mini-boson stars, the ADM mass increases from the Newtonian limit (as $\omega\rightarrow 1$) until the global maximum is attained. But for sufficiently small $f_a$, this is only a local maximum; as the frequency continues to decrease, the mass has a non-monotonic behaviour, and the global maximum of the mass is attained at the lower end of the frequency interval. Similar trends occur for the angular momentum. This sort of non-spiralling curves, with the maximal mass appearing at the lower end of the frequency interval, can be seen for boson stars with a $Q$-ball (sextic) potential, wherein the quadratic, quartic and sextic terms alternate in sign~\cite{Kleihaus:2005me,Kleihaus:2007vk}.  

Finally, a curious and intriguing feature in Fig.~\ref{Fig:Mass} is that despite the $f_a$ dependence of the curves, for all $f_a$, the space of solutions always appears to converge to similar values, albeit not exactly the same, of $M,J,\omega$ as one moves along the curve towards the strong gravity region (the "centre" of the spiral).

\begin{figure}[ht!]
\begin{center}
\includegraphics[height=.38\textheight, angle =0]{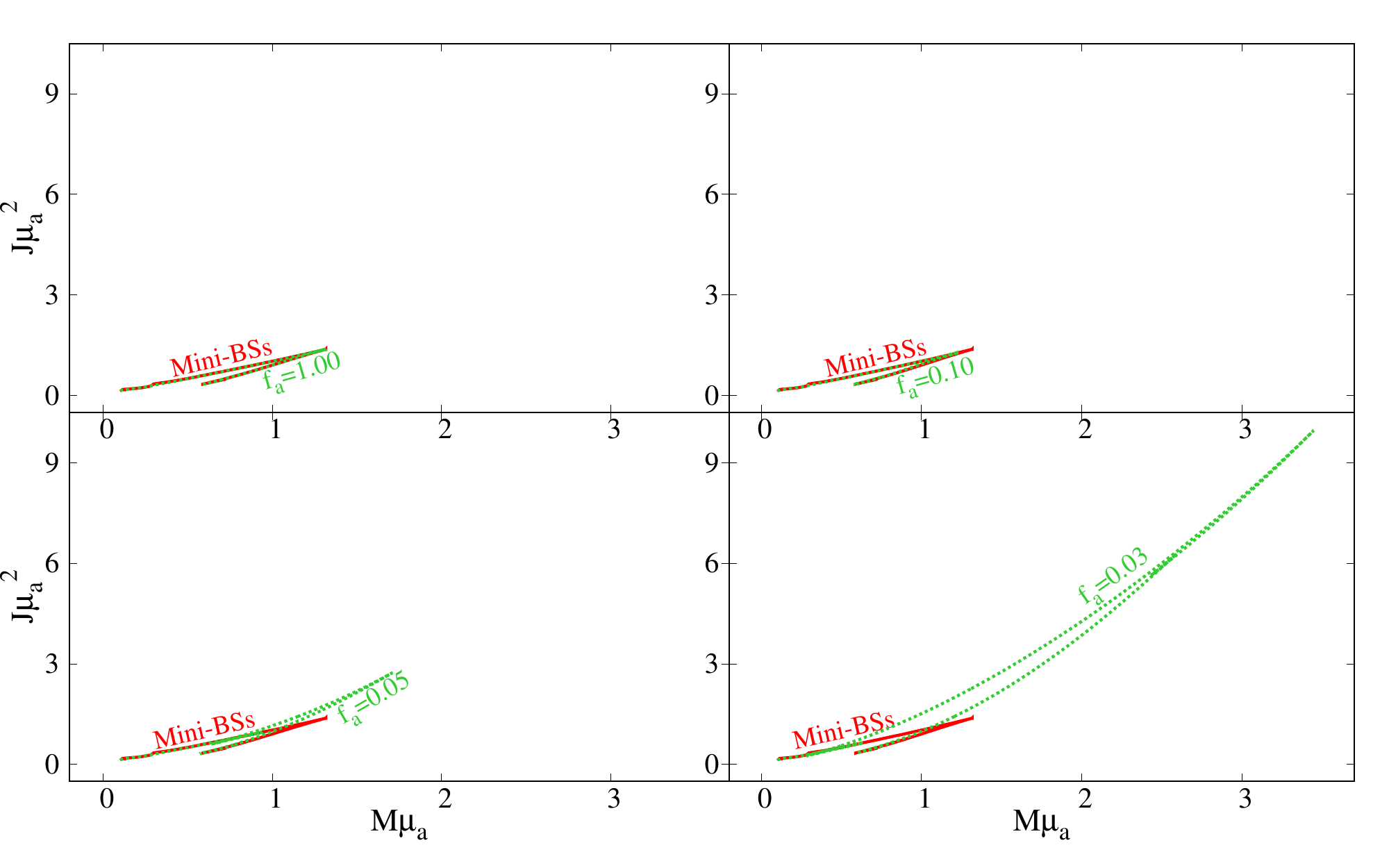} \\  \vspace{-0.5cm}
\includegraphics[height=.38\textheight, angle =0]{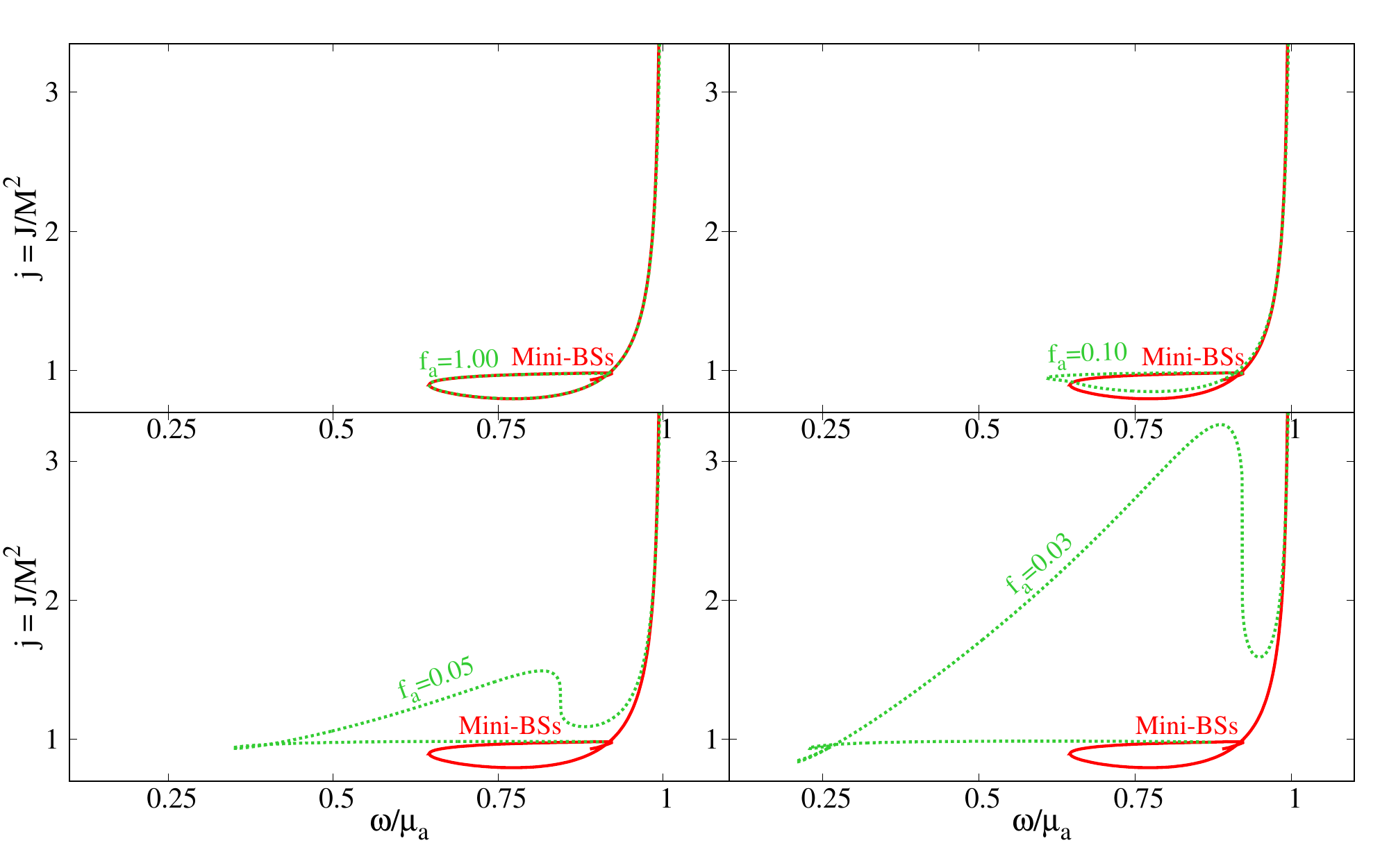}
\end{center}
  \vspace{-1cm}
\caption{(Top panels) Total angular momentum $vs.$ ADM mass. (Bottom panels) Dimensionless spin $j \equiv J/M^2$ $vs.$  the angular frequency $\omega$.}
\label{Fig:MassVsAngMom}

\end{figure}

The bottom panels in Fig.~\ref{Fig:MaxPhi} show the inverse of the compactness, $R_{99}/R_{\text{Schw}}$ $vs.$ the angular frequency. All solutions, regardless of the decay constant $f_a$, have an inverse compactness larger than unity, meaning that they are always less compact than a black hole, as expected. Moreover, decreasing $f_a$, more compact RABSs can be obtained, reaching closer to the BH limit, $R_{99}/R_{\text{Schw}} = 1$. Finally, we remark that, unlike $\phi_{\text{max}}$, the compactness is not a monotonic function along the solutions curve.

Fig.~\ref{Fig:MassVsAngMom} analyses some further features of the total angular momentum, $J$. In the top panels, $J$ is shown against  ADM mass. This phase space confirms that $M,J$ are positively correlated: they either both increase or both decreasing, as one moves along the solution space, regardless of $f_a$. The bottom panels
exhibit how the dimensionless spin $j \equiv J/M^2$ changes with the angular frequency. In the Newtonian limit, RABSs, as most macroscopic objects, can violate the Kerr bound, $j \leqslant 1$. As one moves along the solution space away from this limit, the solutions become more compact and violations of the Kerr bound may cease. As $f_a$ decreases, the Kerr bound abiding solutions seem to become less pronounced in solution space.

\subsection{Other properties}

\subsubsection{Ergoregions}
The existence of an ergoregion is an important property of the Kerr spacetime, with remarkable physical consequences, namely the possibility of energy extraction, as first pointed out by Penrose~\cite{Penrose:1969pc,Penrose:1971uk}. For horizonless ultracompact objects, like spinning boson stars, the presence of an ergoregion means, generically, an instability~\cite{Cardoso:2007az}. Thus, it is relevant to analyse the occurrence of an ergoregion for RABSs. Previous analysis of ergoregions in models of spinning boson stars can be found in, $e.g.$~\cite{Kleihaus:2007vk,Herdeiro:2014jaa,Herdeiro:2018djx,Kunz:2019sgn,Kunz:2019bhm}.

\begin{figure}[ht!]
\begin{center}
\includegraphics[height=.380\textheight, angle =0]{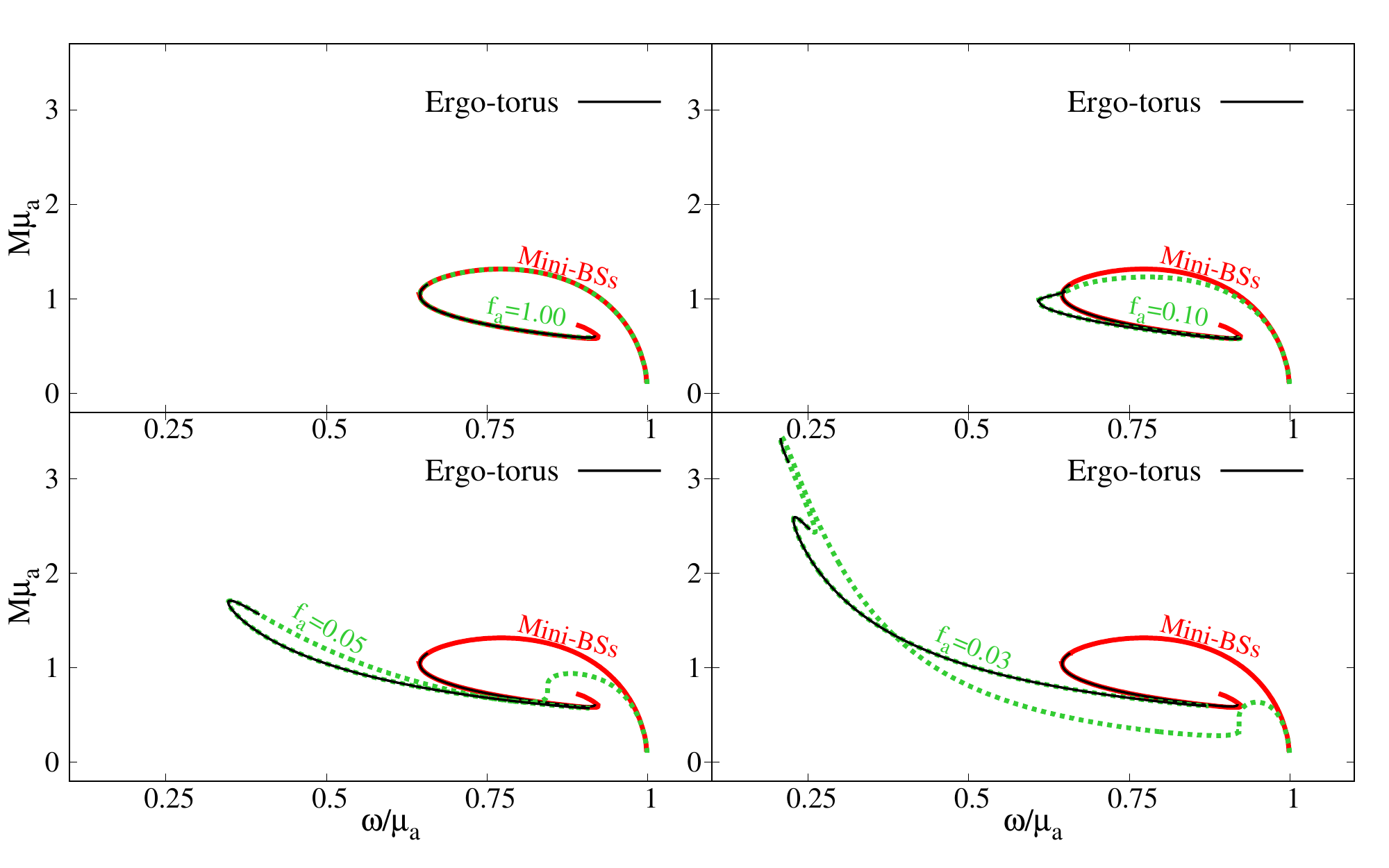}
\end{center}
  \vspace{-0.5cm}
\caption{Occurrence of an ergo-torus in RABSs.}
\label{Fig:Ergoregions}
\end{figure}

In asymptotically flat spacetimes, ergoregions are defined as the spacetime domain in which the norm of $\xi = \partial_t$ becomes positive. An ergoregion is bounded by the surface $\xi^2 = 0$, or in terms of the metric functions,
\begin{equation}
	g_{tt} = -e^{2F_0} + W^2 e^{2F_2} \sin^2 \theta  = 0 \ .
\end{equation}   
Ergoregions for spinning boson stars have a toroidal topology, dubbed ergo-torus~\cite{Herdeiro:2014jaa}. They are not present in the Newtonian limit, as $\omega\rightarrow 1$. For the case of spinning mini-boson stars, the first solution showing an ergo-region occurs has a lower frequency than the maximal mass solution - Fig.~\ref{Fig:Ergoregions}.

For the case of RABSs, we have found that the behaviour is analogous to that for mini-boson stars. For large values of $f_a$, the solutions start to develop an ergo-torus around the same value of $\omega$ as in the case of mini boson stars. Following the solutions line towards the strong gravity regime, they have always an ergo-torus. Decreasing $f_a$, the first solutions that exhibit an ergo-torus occur for smaller values of $\omega$. This behaviour is shown in Fig. \ref{Fig:Ergoregions}, wherein the part of the solution line that contains ergo-regions is shown as a black solid line. A qualitative novelty is that for sufficiently small values of $f_a$, it is possible to have two disconnected regions where the solutions have ergo-torus. This behaviour is seen for the case with $f_a = 0.03$ in Fig. \ref{Fig:Ergoregions}, and is distinct from the remaining lines. As a quantitative reference, in Table~\ref{Tab:ErgoRegionValues} we provide the data of the first solution exhibiting an ergo-region, when moving along the solution line starting from the Newtonian limit.
%
%
\begin{table}[ht!]
	\centering
	\begin{tabular}{|c|c|c||c|c|}
		\hline
  	\multirow{2}{*}{$f_a$} &\multicolumn{2}{c||}{Ergo-region} &\multicolumn{2}{c|}{LR} \\ 
 		\cline{2-3} \cline{4-5}
		 & $\omega/\mu_a$ & $M\mu_a$ & $\omega/\mu_a$ & $M\mu_a$ \\
		\hline
		\hline
		1.00 & 0.658 & 1.154  & 0.747 & 1.308\\
		0.10 & 0.651 & 1.078 & 0.747 & 1.224 \\
		0.05 & 0.396 & 1.563 & 0.646 & 0.843 \\
		0.03 & 0.207 & 3.428 & 0.336 & 1.712 \\
		\hline
	\end{tabular}
	\caption{Data of the first RABS solution exhibiting an ergo-region or a LR, for several decay constants $f_a$.}
	\label{Tab:ErgoRegionValues}
\end{table}

\subsubsection{Light rings and innermost stable circular orbits}
Besides ergoregions, other two strong gravity features found for geodesic motion around Kerr black holes are the existence of an innermost stable circular orbit (ISCO) for timelike geodesics, and light rings (LRs), for null geodesics~\cite{Bardeen:1972fi}. Again, these features can be found for some, but not all, boson stars.  For instance, spherical mini boson stars do not have an ISCO; but spinning mini boson stars can have, see $e.g.$~\cite{Cao:2016zbh}. On the other hand, boson stars, both static and spinning, only have LRs for sufficiently strong gravity configurations~\cite{Cunha:2015yba} - see also~\cite{Grandclement:2016eng}. Such ultracompact boson stars are perturbatively unstable~\cite{Cunha:2017wao}. Moreover, when LRs are present for boson stars, they always come in pairs~\cite{Cunha:2017qtt}.\footnote{The pair refers to one saddle point and one minimum of the effective potential~\cite{Cunha:2017qtt}. Typically they are both co-rotating or counter-rotating. There may, or may not exist another pair, in the opposite sense of rotation.} Also, the interplay between LRs and the ergoregion can lead to interesting lensing features~\cite{Cunha:2016bjh}.

RABSs with large $f_a$ are identical to mini boson stars; thus they must have an ISCOs and LRs. We shall now analyse how these features change when varying $f_a$. We will follow the standard method to obtain such orbits.

The effective Lagrangian for equatorial, $\theta = \pi/2$, geodesic motion on the geometry \eqref{Eq:AnsatzMetric} is
\begin{equation}
	2\mathcal{L} = e^{2F_1}\dot{r}^2 + e^{2F_2} r^2 \left( \dot{\varphi} - \frac{W}{r} \dot{t} \right)^2 - e^{2F_0} \dot{t}^2 \equiv \epsilon  \ ,
\label{el}
\end{equation}
where all \textit{ansatz} functions depend only on the radial coordinate $r$, the dot denotes derivative \textit{w.r.t.} the proper time and $\epsilon = \{-1,0\}$ for timelike and lightlike particles, respectively. Due to the isometries, we can write $\dot{t}$ and $\dot{\varphi}$ in terms of the energy $E$ and angular momentum $L$ of the test particle,
\begin{eqnarray}
	&&E = \left( e^{2F_0} - e^{2F_2} W^2 \right) \dot{t} + e^{2F_2} r W \dot{\varphi} \ , \\
	&&L = e^{2F_2} r^2 \left( \dot{\varphi} - \frac{W}{r} \dot{t} \right) \ .
\end{eqnarray}
The effective Lagrangian \eqref{el} then yields an equation for $\dot{r}$, which defines a potential $V(r)$,
\begin{equation}
	\dot{r}^2 = V(r) \equiv e^{-2F_1} \left[ \epsilon - e^{-2F_2} \frac{L^2}{r^2} + e^{-2F_0} \left( E - L \frac{W}{r} \right)^2 \right] \ .
\end{equation}
For circular orbits, both the potential and its first derivative must be zero, $V(r) = 0$ and $V(r)' = 0$. For lightlike particles, the former equation gives two algebraic equations, defining two possible impact parameters for the particle, $b_+ \equiv L_+/E_+$ and $b_- \equiv L_-/E_-$, corresponding to co and counter rotating orbits, respectively; the latter equation, together with the impact parameters, yields the two radial coordinates of the co- and counter-rotating LRs. Whenever it is possible to obtain a real solution for the radial coordinate, the RABS possesses LRs. 

\begin{figure}[ht!]
\begin{center}
\includegraphics[height=.380\textheight, angle =0]{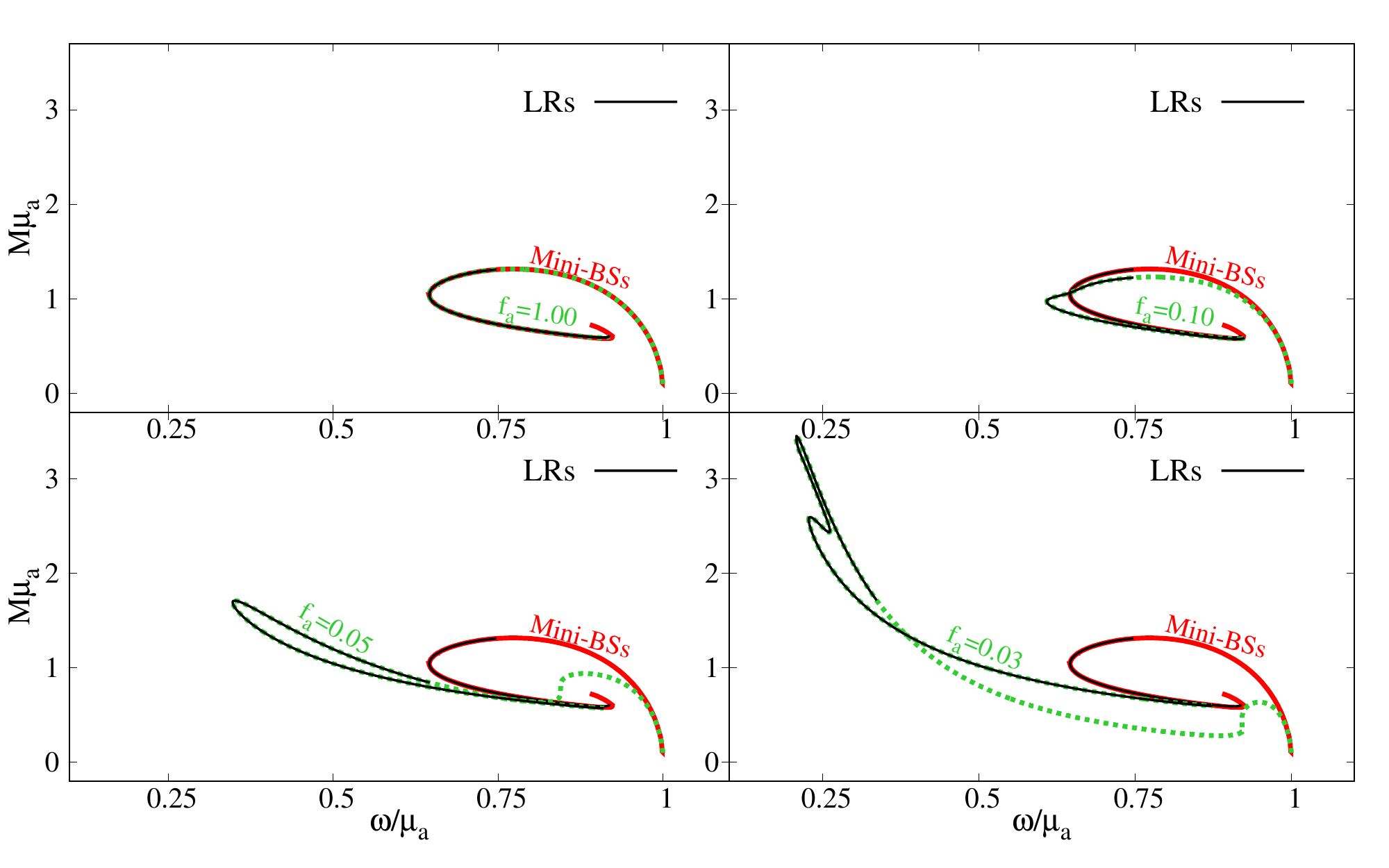}
\end{center}
  \vspace{-0.5cm}
\caption{Occurrence of LRs in RABSs.}
\label{Fig:LR}
\end{figure}

For all RABSs obtained in this work, no solution possesses a co-rotating LR. Counter-rotating LRs, on the other hand, exist for sufficiently strong gravity configurations. This is shown in Fig. \ref{Fig:LR}. For the largest value of $f_a$, LRs start to occur for roughly the same $\omega$ as for mini boson stars~\cite{Cunha:2015yba}. Moving along the solution line, after LRs first occur, they are always present along the solutions line (shown as the black solid line). Decreasing $f_a$, varies the frequency of the solution for which LRs first occur. As a quantitative reference, in Table~\ref{Tab:ErgoRegionValues} we provide the data of the first solution exhibiting a LR, when moving along the solution line starting from the Newtonian limit.

Now we turn to timelike orbits. In this case, the vanishing of the potential $V(r)$ and of its first derivative yields two algebraic equations for the energy and angular momentum of the particle, $\{E_+,L_+\}$ and $\{E_-,L_-\}$ corresponding to co- and counter-rotating orbits, respectively. The stability of such orbits is determined by the sign of the second derivative of the potential $V(r)$. A negative (positive) sign corresponds to stable (unstable) orbits. These are denoted as SCOs and UCOs, respectively.  

\begin{figure}[ht!]
\begin{center}
\includegraphics[height=.380\textheight, angle =0]{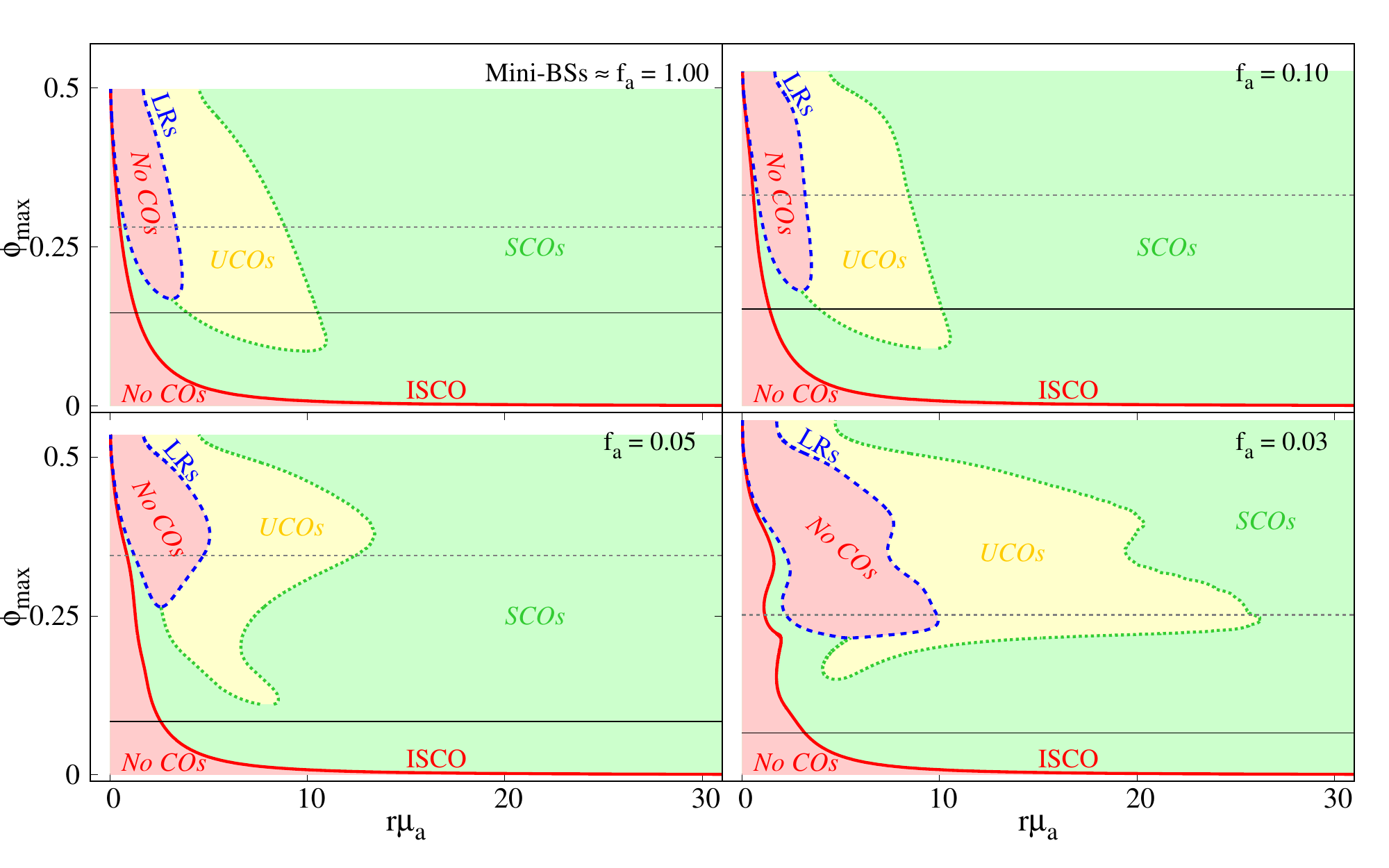}
\includegraphics[height=.380\textheight, angle =0]{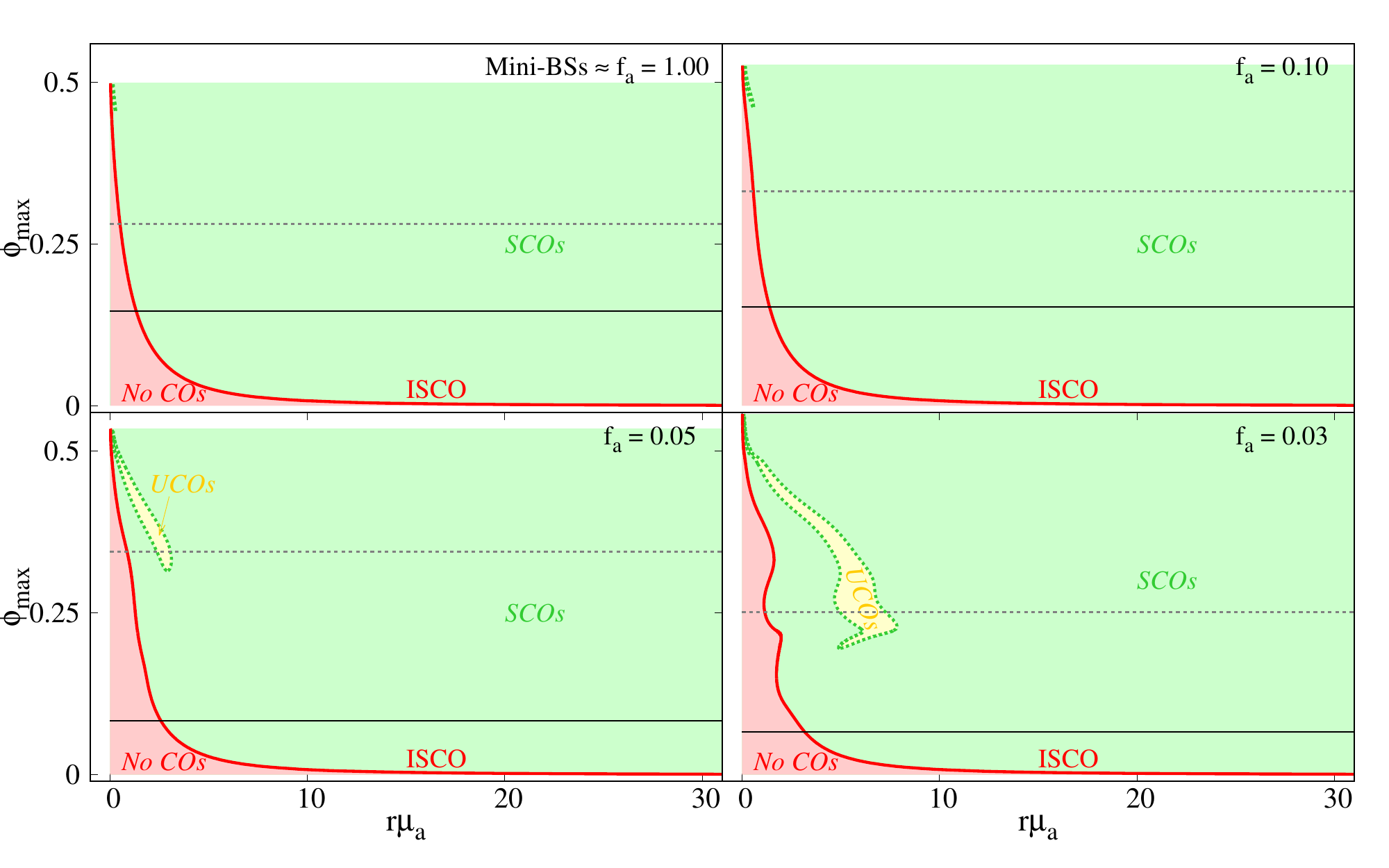}
\end{center}
  \vspace{-0.5cm}
\caption{Structure of counter-rotating (top panels) and co-rotating (bottom panels) circular orbits for the case of mini-boson stars and RABSs with $f_a = \{0.10, 0.05, 0.03\}$. The light (yellow) green region corresponds to timelike SCOs (UCOs). The light red region corresponds to no circular orbits (No COs). The ISCO is represented as a solid red line and the LRs are represented as a dashed blue line. The horizontal black line corresponds to the solution wherein, moving along the line of solutions from the Newtonian limit, the first local maximum of the mass is attained. The horizontal grey dotted lines correspond to the first solution which develops an ergo-torus.}
\label{Fig:ISCO_Counter}
\end{figure}

By studying the stability of the equatorial timelike circular orbits, we found a quite different structure than the one which is commonly found for black holes, and in particular for the Kerr solution. This different structure is exhibited in Fig.~\ref{Fig:ISCO_Counter}, for mini boson stars and RABSs with $f_a = \{0.10, 0.05, 0.03\}$. The mini boson star case is essentially identical to the RABS case with $f_a=1.00$. But since no such plot has been exhibited in the literature for the paradigmatic case of spinning mini boson stars we emphasise it here. 

In Fig.~\ref{Fig:ISCO_Counter} we have chosen to parameterise the RABSs along the line of solutions, for any $f_a$ by the maximum value of the scalar field, since this is a monotonic quantity along this line, $cf.$ Fig.~\ref{Fig:MaxPhi} (top panels): $\phi_{max}$ increases monotonically from the dilute/Newtonian regime ($\omega \sim 1$ and $\phi_{max} \sim 0$) until the strong gravity regime. Therefore, any horizontal line corresponds to a unique RABS solution. 

Fig.~\ref{Fig:ISCO_Counter} exhibits the ISCO, as a red solid line; the LRs, as a blue dashed line; the region of timelike SCOs is filled by a light green colour; the region of timelike UCOs is filled by a light yellow colour; and the region where no timelike circular orbits exist (No COs) is filled by a light red colour. The black solid horizontal line provides a reference in the space of solutions, denoting the solution where the first local maximum of the ADM mass occurs. The grey dotted horizontal line corresponds to the first solution to develop an ergo-torus.

Let us focus first on the counter-rotating orbits -- Fig. \ref{Fig:ISCO_Counter} (top panels). 
Dilute solutions ($\phi_{max} \sim 0$) only have two different regions: for sufficient small radii it is impossible to have circular orbits since the energy and angular momentum of the particle become imaginary (No COs). But after some radius, it possible to have circular orbits, and they are stable (SCOs). The two regions are separated by the ISCO. 
For less dilute solutions ($\phi_{max} \sim 0.12$, for mini boson stars), a new region with UCOs emerges. This region occurs between a minimum and maximum radius and it is not connected with the ISCO: there are still SCOs between the ISCO and the region with UCOs. This is a Kerr-unlike feature and may be associated with the toroidal structure of spinning boson stars.
Moving further into the strong gravity region LRs emerge, as a pair, delimiting a new region with no COs between them. This new region is connected to UCOs for large radii and to SCOs for small radii. Moving further into the strong gravity regime, at larger values of $\phi_{max}$,  the two regions with no COs appear to converge, but there is always a small region of SCOs  for all solutions we obtained.

All features described in the previous paragraph apply to both to mini boson stars and RABSs with different $f_a$. The main difference resides in the size of the different regions. Decreasing $f_a$, the regions of no COs and UCOs get broader and their shapes change, with an increased complexity associated with the several back-bendings that appear for solutions with smaller $f_a$ -- see \textit{e.g.} Fig. \ref{Fig:Mass}. Another difference is related to the solution with the first local maximum of the mass (horizontal black line in Fig. \ref{Fig:ISCO_Counter}). For sufficiently small values this solution has no UCOs.

The case of co-rotating orbits -- Fig. \ref{Fig:ISCO_Counter}  (bottom panels) is much simpler. The key features are the following. Firstly, as mentioned above, co-rotating LRs do not exist for RABSs. Secondly, albeit hardly noticeable for mini-boson stars and RABSs with $f_a = 0.10$, there is always a region of UCOs for the solutions in the strong gravity regime. This region of UCOs starts to get broader and more complex when moving into the strong gravity regime, analogously to the counter-rotating case. Thirdly, all solutions below the horizontal black lines have the same structure, meaning that they only have a region of no COs and another of SCOs.

\subsubsection{Energy conditions}
Possible violations of energy conditions are informative about exotic properties of the matter-energy content necessary to support some spacetime configuration. Let us briefly analyse the weak, dominant and strong energy condition, WEC, DEC and SEC, respectively, for the axion-like potential model and RABS solutions considered in this paper. 

The WEC is defined as the requirement that $T_{\mu\nu} X^\mu  X^\nu \geqslant 0$, for any  timelike vector field $X^\mu$. It means the energy density measured by any timelike observer must  non-negative. The DEC amounts to the WEC plus the requirement that for every future directed causal vector field $X^\mu$, the vector $T_{\mu\nu} X^\nu$ is causal and future directed. This extra requirement means that the energy flux is causal (timelike or null). The SEC is defined as $\left( T_{\mu\nu} - \frac{1}{2} g_{\mu\nu} T \right) X^\mu  X^\nu \geqslant 0$, and it means matter gravitates towards matter. Let us test these conditions for RABSs. Previous considerations on energy conditions for spinning boson stars can be found, $e.g.$ in~\cite{Collodel:2019ohy}.

Consider a generic unit timelike vector, $X^\mu X_\mu = -1$. Then, for our model:
\begin{equation}
	T_{\mu\nu} X^\mu X^\nu = 2 X^\mu \partial_\mu \Psi^* X^\nu \partial_\nu \Psi + \partial^\alpha \Psi^* \partial_\alpha \Psi + V \ .
\end{equation}
Since the three terms on the right-hand side of the above equation are non-negative, their sum will be non-negative. Therefore the WEC is never violated. To analyse the DEC, we compute the norm of the vector $T_{\mu\nu} X^\nu$, which is,
\begin{equation}
	||T_{\mu\nu} X^\nu||^2 = 2 \partial_\mu \Psi^* \partial^\mu \Psi^* (X^\nu \partial_\nu \Psi)^2 - 2 X^\mu \partial_\mu \Psi^* X^\nu \partial_\nu \Psi \left( \partial^\alpha \Psi^* \partial_\alpha \Psi + 2 V \right) - \left( \partial^\alpha \Psi^* \partial_\alpha \Psi + V \right)^2.
\end{equation}
This expression is not manifestly sign-definite. But specialising it for the \textit{ansatz} used in this work, we obtain,
\begin{equation}
	||T_{\mu\nu} X^\nu||^2 = -\frac{4}{r^2} e^{-2(F_0+F_1)}\left(\omega - m \frac{W}{r} \right)^2 \left[ (\partial_\theta \phi)^2 + r^2 \left( e^{2F_1} V + (\partial_r \phi)^2 \right) \right] - \left( \partial^\alpha \Psi^* \partial_\alpha \Psi + V \right)^2 \ .
\end{equation}
The norm of $T_{\mu\nu} X^\nu$ is now manifestly non-positive; thus the energy flux is timelike or null and consequently the DEC is obeyed. To analyse the SEC we consider
\begin{equation}
	\left( T_{\mu\nu} - \frac{1}{2} g_{\mu\nu} T \right) X^\mu  X^\nu = 2 X^\mu \partial_\mu \Psi^* X^\nu \partial_\nu \Psi - V    .
\end{equation}
Since the potential $V$ is always non-negative, the SEC is violated if the potential is large enough. It is well known that the SEC can be violated by 
 a free massive scalar field - see $e.g.$ \cite{Visser:1995cc}. Here we show that such violation is also possible for the axion-like complex scalar field.

\begin{figure}[ht!]
\begin{center}
\includegraphics[height=.280\textheight, angle =0]{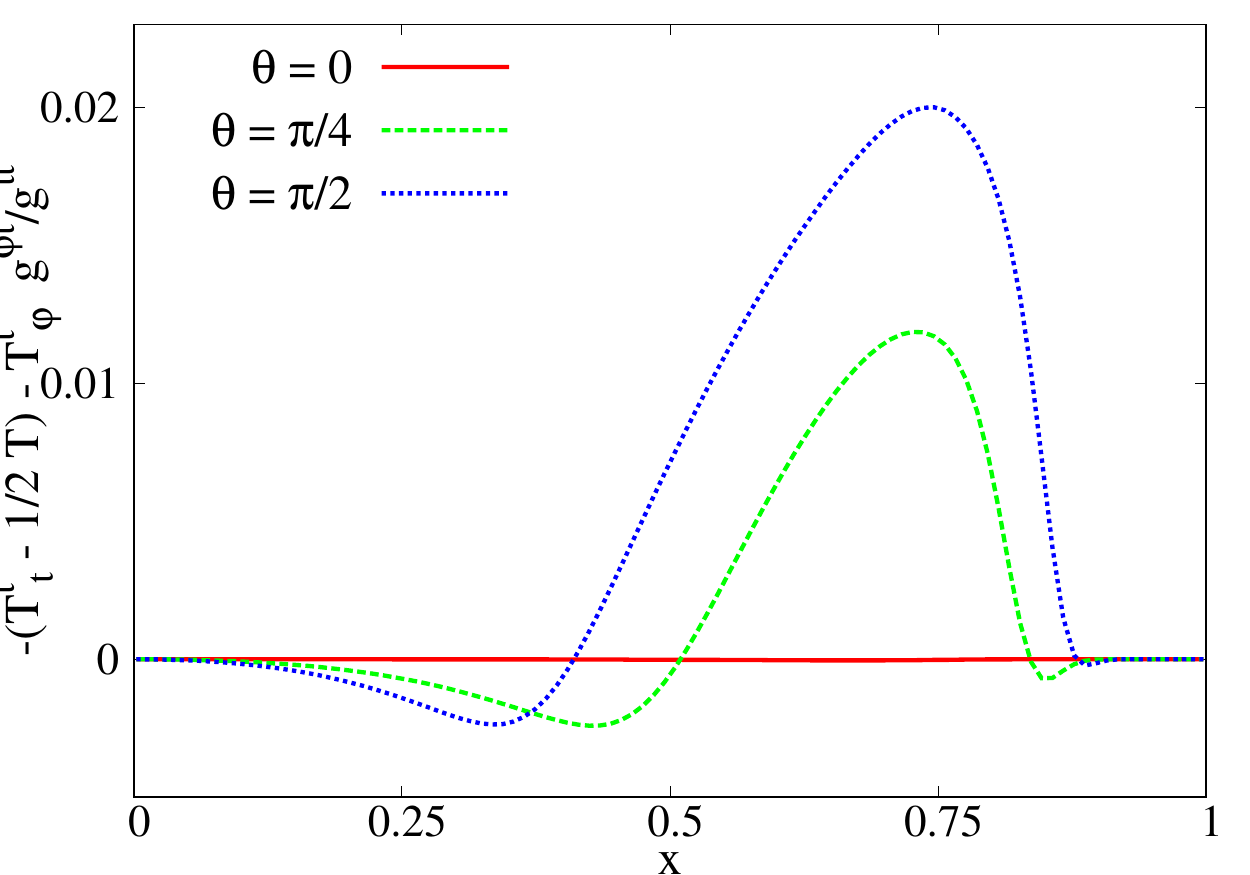}
\end{center}
  \vspace{-0.5cm}
\caption{SEC as a function of the compact radial coordinate $x = r/(1+r)$, for three different polar angles $\theta$, of a ABS solution with $f_a = 0.030$ and $\omega = 0.250$ on the first branch.}
\label{Fig:SEC}
\end{figure}

As an example to study both energy conditions, consider an observer whose 4-velocity is orthogonal to $t = const$ hypersurfaces. Such observers are commonly called Zero Angular Momentum Observers (ZAMO). Their 4-velocity is defined as $u_\mu = (u_t,0,0,0)$. For a ZAMO, the energy conditions are,\footnote{We do not present the inequality associated to the DEC which is too complicated to display here.}
\begin{eqnarray}
	&&\text{WEC:} \hspace{10pt} - T^t_t - T^t_\varphi \frac{g^{\varphi t}}{g^{tt}} \geqslant 0 \ . \\
	&&\text{SEC:} \hspace{10pt} - \left(T^t_t - \frac{1}{2} T\right) - T^t_\varphi \frac{g^{\varphi t}}{g^{tt}} \geqslant 0 \ .
\end{eqnarray}
We found that, for a ZAMO, the WEC and DEC is never violated, as shown in general in the previous paragraph; but the SEC is violated for some solutions with small values of $f_a$. This is illustrated in Fig. \ref{Fig:SEC}. Therein, a RABS solution, with $f_a = 0.030$ and $\omega = 0.250$ belonging to the first branch (top branch) is chosen, and one can see it has some space-time regions wherein the SEC is violated. This illustration raises two questions: $i)$ for which values of $f_a$ is it possible to obtain solutions that violate the SEC? $(ii)$ For each line of constant $f_a$ that has solutions that violate the SEC, where do these solutions occur and how do they depend on $f_a$?

\begin{figure}[ht!]
\begin{center}
\includegraphics[height=.380\textheight, angle =0]{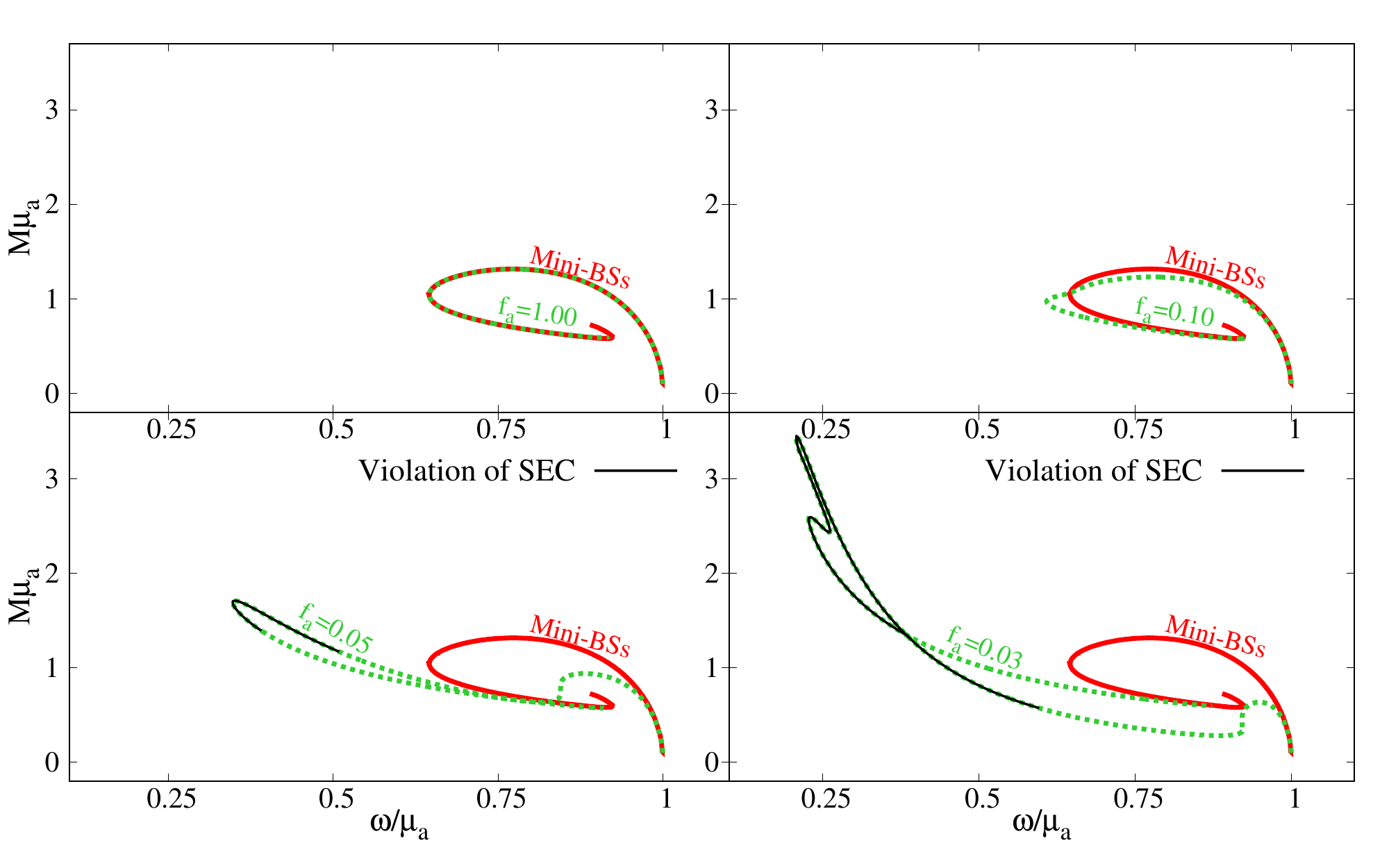}
\end{center}
  \vspace{-0.5cm}
\caption{SEC violations in RABSs.}
\label{Fig:ViolationSEC}
\end{figure}

As for the first question, we found that solutions that violate the SEC start to appear for $f_a \lesssim 0.057$, on a small region of solutions close to the end of the first branch (close to the first back bending). The second question is addressed in Fig. \ref{Fig:ViolationSEC}. One can observe that for the line with constant $f_a = 0.050$, there are solutions in the second branch that violate the SEC. By decreasing further $f_a$, the region of solutions that violate the SEC gets broader, as shown in the bottom right panel ($f_a = 0.030$). Further decreasing $f_a$, one may extrapolate that, for small enough $f_a$, a large portion of the solutions with that $f_a$ may violate the SEC. We emphasise this analysis was performed assuming a physical ZAMO observer.

\section{Conclusions and remarks}

In this work we have constructed RABSs, which are the spinning generalisation of static axion boson stars recently found in~\cite{Guerra:2019srj}. These are stationary, axially symmetric, asymptotically flat and everywhere regular solutions of the Einstein-Klein-Gordon theory with a QCD axion potential defined in Eq. \eqref{Eq:Potential}. The resulting solutions are described by two parameters: the angular frequency of the scalar field $\omega$ and the decay constant of the QCD potential $f_a$. For large $f_a$, the solutions reduce to the standard mini boson stars. In practice, $f_a = 1.00$ is already large.

By comparing RABS with the limiting case of mini boson stars, we observed that the solutions with the smaller values of the decay constant can possess lower values of the angular frequency of the scalar field, be more massive and have larger values for angular momentum and scalar field, leading to more compactness configurations. At the level of the phenomenology, both RABSs can develop LRs if they are sufficiently in the strong gravity part of the space of solutions. The structure of timelike, equatorial circular orbits is also similar for RABSs and mini boson stars; but for low values of $f_a$ the structure starts to be more complex. An important difference arises at the level of the strong energy condition. For a ZAMO, mini boson stars never violate the strong energy condition, but for RABS they can violate that energy condition if their decay constant is low enough.

An interesting direction to follow up on this work is to consider a small value of the decay constant, \textit{e.g.} $f_a = 0.03$, and add a black hole horizon in the middle of the RABS, under a synchronisation condition, as in~\cite{Herdeiro:2014goa}.  This will lead to black holes with axionic hair. The phenomenological properties of such black holes could be compared to current observations as a case study to test the no-hair hypothesis. Finally, it would be interesting to study the stability of these solutions using fully non-linear numerical relativity simulations, following~\cite{Sanchis-Gual:2019ljs}, to assess the impact of varying $f_a$ on the development of the non-axisymmetric instabilities observed for spinning mini-boson stars.

\section*{Acknowledgements}

J. D. is supported by the FCT grant SFRH/BD/130784/2017. This work is supported by the Center for Research and Development in Mathematics and Applications (CIDMA) through the Portuguese Foundation for Science and Technology (FCT - Fundacao para a Ci\^encia e a Tecnologia), references UIDB/04106/2020 and UIDP/04106/2020 and by national funds (OE), through FCT, I.P., in the scope of the framework contract foreseen in the numbers 4, 5 and 6 of the article 23, of the Decree-Law 57/2016, of August 29, changed by Law 57/2017, of July 19. We acknowledge support from the projects PTDC/FIS-OUT/28407/2017 and CERN/FIS-PAR/0027/2019. This work has further been supported by the European Union's Horizon 2020 research and innovation (RISE) programme H2020-MSCA-RISE-2017 Grant No.~FunFiCO-777740. The authors would like to acknowledge networking support by the COST Action CA16104.

\bibliographystyle{JHEP}
\bibliography{References}

\end{document}